\documentclass[a4paper, preprint, preprintnumbers, superscriptaddress, showpacs, showkeys, nofootinbib, floatfix, twocolumn, aps, pra, 10pt]{revtex4-2}
\usepackage[utf8]{inputenc}
\usepackage[english]{babel}
\usepackage[dvips]{graphicx}
\usepackage[breaklinks=true,colorlinks,citecolor=blue,linkcolor=blue,urlcolor=blue]{hyperref}
\usepackage{xcolor}
\usepackage{url}
\usepackage{amsthm,amsmath,amssymb,bbm,mathtools}
\usepackage{physics}
\usepackage{appendix}

\usepackage{caption}
\usepackage{subcaption}
\newenvironment{subsubfigure}[2][]{%
  \begin{subfigure}[#1]{#2}%
    \stepcounter{subsubfigure}%
}{%
    \addtocounter{subfigure}{-1}%
  \end{subfigure}%
}

\newcounter{subsubfigure}

\usepackage{orcidlink}

\allowdisplaybreaks

\begin{document}

\title[LAQC RTN]{Quantum correlations in mutually unbiased bases and non-Markovian dephasing}

\author{Hermann L. Albrecht\,\orcidlink{0000-0002-5735-8340}}
\email{albrecht@usb.ve}
\affiliation{Departamento de F\'{\i}sica, Universidad Sim\'on Bol\'{\i}var, AP 89000, Caracas 1080, Venezuela.}

\author{David M. Bellorin\,\orcidlink{0009-0008-7881-8095}}
\email{dmbellorin@usb.ve}
\affiliation{Departamento de F\'{\i}sica, Universidad Sim\'on Bol\'{\i}var, AP 89000, Caracas 1080, Venezuela.}
\affiliation{Laboratorio de Sistemas Complejos, Universidad Sim\'on Bol\'{\i}var, AP 89000, Caracas 1080-A, Venezuela.}

\date{August 12, 2026}

\begin{abstract}
Wu et al. introduced a symmetric measure $Q^{s}$ in terms of mutually unbiased bases (MUBs) that quantifies residual quantum correlations (RQC). The definition of this measure involves determining two complementary local bases via an optimization of the quantum mutual information. One defines the first local bases, the optimal computational ones, by maximizing the classical correlations measure. Next, one defines a new set of local bases that are complementary to the first ones. In the latter, one also calculates the corresponding measure of quantum correlations by maximization. Considering alternative optimizations of the mentioned measures, one can define other quantifiers. In particular, Mundarain and Ladrón de Guevara introduced local available quantum correlations (LAQC, $\mathcal{L}$) in terms of the complementary local bases to the ones defining the least classically correlated state. That is, they characterize LAQC by defining the optimal bases as those that minimize the classical correlations measure. This type of quantum correlation is closely related to an entanglement activation protocol proposed by Piani et al. In previous articles, we derived an exact solution for the LAQC measure for 2-qubit X states. Based on these previous results and methodology, we determine the corresponding exact analytical expression for $Q^s$ of 2-qubit X states. We then analyze the behavior of these two measures for two non-Markovian quantum dephasing channels: Random Telegraph (RT) and Modified Ornstein–Uhlenbeck (MOU) noises. We derive general conditions for sudden death and revival of these measures in X states and illustrate these results with four families of bipartite qubit states: Werner states, Maximally Nonlocal Mixed States (MNMS), Maximally Entangled Mixed States (MEMS), and a mixed symmetric separable state.
\end{abstract}
\preprint{SB/F/498-25}
\maketitle


\section{Introduction}\label{sec:Intro}

The interaction of a system with its environment is unavoidable in real-world scenarios where we intend to implement quantum information tasks \cite{Nielsen-QIT, Nakahara-QC, Vedral-QIT}. Said interaction mediates the loss of coherence and is detrimental to quantum correlations, the core resource in quantum information science and technology. Therefore, the study and analysis of quantum open systems has become an intense field of research \cite{Aolita2015OpenSysDyn-Review, Schlosshauer2019QuantumDeco}.

The study of such systems oftentimes relies on the use of a master equation \cite{Schlosshauer, Campaioli2024QuantumMasterEq} derived from the Liouville-von Neumann equation for the system-environment state. When the Markov and Born approximations apply, these master equations are known as the Gorini–Kossakowski–Sudarshan equation \cite{Gorini1976-GKS_Eq}, which can be further simplified into a Lindbladian form \cite{Lindblad1976-MasterEq}. Although often called the Lindblad equation, it is also commonly referred to as the Gorini–Kossakowski–Sudarshan-Linblad (GKSL) equation \cite{Manzano2020-GKSL_Eq}.

The quantum dynamical semigroup underlying the Lindblad equation allows one to express this time evolution as a map analogous to a superposition of transformations with a finite set of operators that obey a closure relation \cite{Havel2003-KrausLind, Nakazato2006-KrausLind, Andersson2007-KrausLind}. Such operators are labeled as the Kraus operators \cite{Kraus-Article, Kraus-Lectures} of the quantum channel.

On the other hand, analyzing dynamics where the Markovian approximation is no longer valid broadens the type of physical processes that can be studied. Although many publications have studied such dynamics via a master equation \cite{GardinerZoller-QuantumNoise}, in some cases, one can analyze non-Markovian quantum decoherence with an operator-based approach, analogous to the Kraus formalism. In particular, some proposals for non-Markovian quantum noise rely on the extension of the Kraus operators for Markovian quantum channels, such as the dephasing (phase damping) or phase-flipping channels \cite{Daffer2004-QuantumRTN, Kumar2018-NonMarkovianGameTheory, Thapliyal2017-qCrypto-NonMarkovChannel, Naikoo2019-QI_NonMarkovEvol, Mishra2022-QIP-NonMark-Xstates, Sabale2024Dchannels-RTN-OUN, Aiache2025-Dephasing}. 

The memory kernel present in these non-Markovian master equations generally allows one to model colored noise  \cite{GardinerZoller-QuantumNoise, Benedetti2013-ColoredNoise, Benedetti2014-RTN&ColoredNoise}. Among these type of noises, several authors have analyzed Random Telegraph Noise \cite{Daffer2004-QuantumRTN} (RTN) and the Modified Ornstein–Uhlenbeck Noise \cite{Uhlenbeck1930O-OU-original, Maller2009-Gen_OrnsteinUhlenbeck} (MOUN).

Studying the dynamics of multipartite open systems can become a challenging task for many-partite ones, where the complexity can increase dramatically with the number of subsystems involved. Nevertheless, there have been results that point to a universal behavior for the entanglement dynamics in the large-size limit of many-particle systems \cite{Tiersch2013Universality}. Practical applications in quantum information science and technology require the theoretical understanding of such large systems.

Despite this, a route often taken in this endeavor is to understand much simpler systems. In this sense, the simplest multipartite quantum system is a bipartite qubit system, and the so-called X states \cite{EstadosX}
\begin{eqnarray}\label{eq:EstadosX-rho_ij}
\rho_X =\mqty(
   \rho_{11} & 0 & 0 & \rho_{14} \\
   0 & \rho_{22} & \rho_{23} & 0 \\
   0 & \rho_{32} & \rho_{33}& 0 \\
   \rho_{41}& 0 & 0 & \rho_{44})
\end{eqnarray} 
have caught special attention within the quantum information community. These states naturally appear in several theoretical and experimental settings \cite{Quesada-XStates},  are interesting from a mathematical standpoint \cite{Rau_2009-Xstates_algebra}, and a large number of important entanglement measures are exactly computable. They are also of special interest since they constitute a representative class of 2-qubits  \cite{XStates-Entanglement, Hedemann-XStates}.

Moreover, instead of the seven independent real parameters required to characterize $\rho_X$ \eqref{eq:EstadosX-rho_ij}, Zhou et al. \cite{Zhou-CanonicalXstates} established that for $\rho_{14}, \rho_{23} \in \mathbbm{R}$, $\rho_X$ still remains as the representative class, reducing to only five the number of required independent real parameters. Therefore, we can rewrite the above density matrix as
\begin{eqnarray}\label{eq:estadosX-real}
\rho_X =\mqty(
   a & 0 & 0 & r \\
   0 & b & s & 0 \\
   0 & s & c& 0 \\
   r& 0 & 0 & d)
\end{eqnarray}
where
\begin{subequations}
\begin{align}
    &a,b,c,d,r,s\in \mathbbm{R},\\
    &a,b,c,d\geq0,\\
    &|s|\leq\sqrt{bc},\qq{and}|r|\leq\sqrt{ad},
\end{align}
\end{subequations}
so that $\rho_X$ is hermitian and positive semi-definite. Imposing $a+b+c+d=1$ guarantees the normalization of the state.

As an entanglement measure, Concurrence was introduced by Wooters \cite{Wooters_Concurrence} in relation to the Entanglement of Formation. It is defined as
\begin{equation}\label{eq:DefConcurrencia}
\mathcal{C}\qty(\rho_{AB}) \equiv \max\qty{0, \sqrt{\lambda_1\,} -\sqrt{\lambda_2\,} -\sqrt{\lambda_3\,} -\sqrt{\lambda_4\,}},
\end{equation}
\noindent{}where $\qty{\lambda_i}$ are the decreasing ordered eigenvalues of $\sqrt{\rho_{AB}}\,\tilde{\rho}_{AB}\sqrt{\rho_{AB}}$, with
$\tilde{\rho}_{AB} = (\sigma_y\otimes\sigma_y)\rho_{AB}^*(\sigma_y\otimes\sigma_y)$, $\rho_{AB}^*$ the complex conjugate of $\rho_{AB}$, and $\sigma_y$ is the corresponding Pauli matrix. For X states, a direct calculation leads to the following simple expression  \cite{EstadosX}:
\begin{equation}\label{eq:Concurrencia_X}
\mathcal{C}_X = \frac{1}{2}\,\max \left\{0,\mathcal{C}_1,\mathcal{C}_2\right\},
\end{equation}
\noindent{where}
\begin{equation*}
\mathcal{C}_1\equiv 2\left(|\rho_{14}|-\sqrt{\rho_{22}\,\rho_{33}}\right)\qc \mathcal{C}_2\equiv 2\left(|\rho_{23}|-\sqrt{\rho_{11}\,\rho_{44}}\right).
\end{equation*}

Since quantum correlations are the main resource in quantum information, computation, and communication, it is essential that one understands how different types of quantum noises, whether Markovian or not, affect them. Although a large number of studies have focused on analyzing the dynamics of entanglement \cite{Horodecki-Ent, Tiersch2009EquationEntanglement}, the introduction of quantum discord \cite{qDiscord-Olliver, qDiscord-Henderson} and other quantum correlations \cite{Modi-qDiscord} has expanded the focus on the approach to quantum information resources and features. For instance, Naikoo et al. \cite{Naikoo2019-QI_NonMarkovEvol} studied the time evolution of Quantum Fisher Information \cite{Braunstein1994-qFisherInfo, Braunstein1996-qFisherInfo}, coherence, and purity under non-Markovian dephasing with random telegraph noise. On the other hand, Aiache et al. \cite{Aiache2025-Dephasing} analyzed inherent aspects of non-Markovian phase-flipping channels with colored noise. They focused on the quantumness of the channel \cite{Shahbeigi2018-QuantumnessOfChannels}, the Holevo information \cite{Holevo}, among others, as well as in-state distinguishability.

A rather understudied type of quantum correlations are those defined in terms of complementary local bases \cite{Wu_AMID, Wu_ComplementaryBases}. To the best of our knowledge, the QIST community has mostly overlooked those defined in terms of mutually unbiased bases \cite{Schwinger_MUB, Bengtsson2007-MUB, DURT2010-MUB} (MUBs). In 2015, further developing previous results by Wu and collaborators, Mundarain and Ladrón de Guevara introduced a new MUB-defined quantum correlation (MUB-QC), labeled as local available quantum correlations \cite{LAQC} (LAQC). These are quantum correlations that can be available via local bipartite measurements, which they demonstrated are closely related to the entanglement activation protocol developed by Piani et al. \cite{Piani_2011-EntActiv}. 

Focusing on the bipartite setting, this procedure is able to transform non-classical correlations between two subsystems A and B into distillable entanglement between the subsystems and local ancillas, with experimental setups demonstrating the ability to use quantum discord to perform such a feat \cite{EntAct_Exp_Mazzola2011, EntAct_Exp_Adesso014}. Moreover, methods for utilizing this same protocol to transfer entanglement (via swapping) to the starting subsystems have been studied \cite{EntAct_Swapping_Gharibian2011}.  As entanglement is a verified quantum resource, Piani's protocol could turn these non-classical correlations into precursors of quantum resources, even if they are not proven as such by themselves.  

Although there are still various open questions regarding LAQC, including whether they offer operational advantages or performance gains, in recent years we have derived exact results for its measure for Bell Diagonal States \cite{LAQC_BD, LAQC_BD-Err} as well as for 2-qubit X states \cite{LAQC_Xstates-sym, LAQC_Xstates-no_sym}. We also studied its behavior under Markovian decoherence, focusing on the depolarizing, dephasing, and amplitude-damping channels. Under such decoherent channels, LAQC exhibited an asymptotic death instead of the sudden death observed for entanglement \cite{Almeida2007EnvironmentEntSuddenDeath, EntSuddenDeath}. More recently, Albrecht demonstrated its persistence in a standard quantum correlation swapping scheme \cite{LAQC_Swap}.

To deepen the study of LAQC and related MUB-QC, we analyze the behavior of two quantum correlation measures, $Q^s$ and $\mathcal{L}$, in a specific type of quantum non-Markovian dynamics: colored noise dephasing, particularly the cases of RTN \cite{Daffer2004-QuantumRTN} and MOUN \cite{Maller2009-Gen_OrnsteinUhlenbeck, Kumar2018-NonMarkovianGameTheory}. Given the way these phase-flipping channels affect them, we obtain general results for 2-qubit X states. We further illustrate them by studying four different state families, namely Werner states \cite{Werner}, Maximally Nonlocal Mixed States \cite{Batle2011Nonlocality, Fan2019Inequality} (MNMS), Maximally Entangled Mixed States  \cite{Munro2001Maximizing} (MEMS), and a mixed symmetric separable state \cite{LAQC_Swap}. Contrary to our previous results for Markovian quantum channels \cite{LAQC_BD, LAQC_Xstates-no_sym}, we observe that MUB-QC measures experience sudden death and revival under RTN.

We organized this article as follows. In Section \ref{sec:LAQC}, we provide a brief overview of the definitions of $Q^s$ and LAQC as MUB-defined quantum correlation (MUB-QC) measures and summarize the previously obtained results for the LAQC measure of X states. We apply the same procedure to the symmetric measure $Q^s$, obtaining an expression for X states as well. Then, in Section \ref{sec:QuantumNoise}, we introduce the two non-Markovian noise channels analyzed in this work and present the results of applying them as a common bath to a 2-qubit system in Section \ref{sec:Results}. In Section \ref{sec:Conclusions}, we give a summary of our results and the conclusions of this paper. In a final Appendix, we present the expressions of the diagonal elements of $\rho_X$ in the optimal computational basis.  


\section{Brief recount of $Q^s$ and LAQC as measures of MUB-defined quantum correlations}\label{sec:LAQC}

In \cite{Wu_ComplementaryBases}, Wu et al. focused their analysis on quantum correlations in mutually unbiased bases (MUB-QC), revealing a series of residual quantum correlations (RQC). Although the authors aimed at a discord-like measure of RQC, they defined in a final appendix a symmetric correlation vector, which in turn led them to a measure of classical correlations and the related RQC quantifier.

Given a computational basis $\qty{\ket{i\,j}}$, the probability that the resulting state after a projective measurement is $\ket{i,j}$ is given by $p_{ij}=\Tr\qty(\Pi_i\otimes\Pi_j\,\rho^{AB})$. In this context, the classical mutual information, defined as
\begin{equation}\label{eq:ClassicalMutualInfo}
    I\qty(p_{ij}) = H(p_i^A)+H(p_j^B)-H(p_{ij}),
\end{equation}
where $H(p_{ij})=-\sum_{ij}p_{ij}\log_2p_{ij}$ is the Shannon entropy \cite{Shannon1948Mathematical} and $p_i^I=\sum_jp_{ij}$, for $I=A,B$, characterizes the results of local measurements. Wu et al. defines the symmetric measure of classical correlations $C^s$ as the maximal classical mutual information over all possible projective local measurements,
\begin{equation}\label{eq:Cs-Wu}
    C^s\qty(\rho^{AB}) = \max_{\qty{\Pi^{A}\otimes\Pi^{B}}} I\qty(p_{ij}).
\end{equation}

Since the above-given measure defines a set of computational bases, there is also a corresponding set of complementary bases so that they are mutually unbiased. One can define such bases by considering transformations by the associated complex Hadamard matrix \cite{Bengtsson2007-MUB, DURT2010-MUB}, 
\begin{equation}\label{eq:Hadamard-Gen}
    \mathbbm{H}(\Phi) = \frac{1}{\sqrt{2\,}}\mqty[1 & 1   \\
   \vb{e}^{i\Phi} & -\vb{e}^{i\Phi}]\;\;\in\;U(2),
\end{equation} 
where $0\leq\Phi\leq2\pi$. Finally, Wu et al. defined their symmetric measure of maximal quantum correlation as the maximum classical mutual information over all possible projective local measurements on bases that are mutually unbiased with the ones defining the classical measure. That is,
\begin{equation}\label{eq:Qs-Wu}
    Q^s\qty(\rho^{AB}) = \max_{\qty{\tilde\Pi^{A}\otimes\tilde\Pi^{B}}}\qty[\max_{\qty{\Pi^{A}\otimes\Pi^{B}}} I\qty(p'_{ij})],
\end{equation}
where $\tilde\Pi^{I}$ are the projectors to the complementary basis. The maximization over the projectors defined by Eq. \eqref{eq:Cs-Wu} is only necessary if there is no unique basis maximizing the classical measure.

Inspired by these results, Mundarain and Ladrón de Guevara further explored those RQC revealed in the complementary local bases, introducing what they labeled as local available quantum correlations (LAQC) \cite{LAQC}. Contrary to the measure defined in Eq. \eqref{eq:Cs-Wu}, Mundarain et al. were interested in defining their measures in relation to the least classically correlated basis. Therefore, they define the classical correlations quantifier as the result of minimizing the quantum mutual information of a post-measurement state without readout over all possible von Neumann measurements. That is,
\begin{equation}\label{eq:C(rho)}
    \mathcal{C}\qty(\rho^{AB}) = \min_{\qty{\Pi^{A}\otimes\Pi^{B}}}I\qty(p_{ij}).
\end{equation}

The projectors involved in the above minimization define the so-called optimal computational basis. The original state $\rho^{AB}$ is rewritten in this basis so that the measure for LAQC is defined as the maximal quantum mutual information of a post-measurement state without readout over all possible von Neumann measurements using bases that are complementary to the optimal one. That is,

\begin{equation}    \label{eq:LAQC-definition}
    \mathcal{L}\qty(\rho^{AB}) = \max_{\qty{\tilde\Pi^{A}\otimes\tilde\Pi^{B}}} I\qty(p'_{ij}).
\end{equation}

\subsection{Two MUB-QC measures for 2-qubit X states}\label{sec:LAQC-Xstates}

As previously stated, 2-qubit X states are widely used in the literature. Therefore, obtaining analytical expressions for the measures of different quantum correlations for this particular family of states has been a goal within the quantum information community. 

For many quantum correlations beyond entanglement, such as quantum discord \cite{qDiscord-Olliver, qDiscord-Henderson}, an exact analytical expression is only possible \cite{GirolamiAdesso-QD-Xstates, Huang-QD-Xstates-WorstCaseScenario} for a limited set of states, such as Werner and Bell-Diagonal states \cite{Luo2008QD-BellDiagonal}. For general X states, we have so far approximate analytical expressions \cite{MaldonadoTrapp-QD, Huang-QD-Xstates-WorstCaseScenario, Quesada-XStates, Liao_QD}. For LAQC, on the other hand, we were able to compute an exact analytical result for general 2-qubit states in 2022 \cite{LAQC_Xstates-sym, LAQC_Xstates-no_sym}. 

Since the Pauli matrices, along with the identity, form a basis for the qubit density matrix space, 2-qubit states can be written in the Fano form \cite{Fano1983} as
\begin{equation}\label{eq:2-qubitFanoForm}
    \rho^{AB}=\frac{1}{4}\, \sum_{\mu,\nu=0}^{3}T_{\mu\nu}\sigma_\mu\otimes\sigma_\nu,
\end{equation}
where $T_{\mu\nu}=\Tr\qty[\qty(\sigma_\mu\otimes\sigma_\nu)\rho^{AB}]$ and $\sigma_0=\mathbbm{1}$. For X states \eqref{eq:estadosX-real}, the five non-zero real parameters are $T_{30}, T_{03}, T_{11}, T_{22}$, and $T_{33}$, which are related to the previous ones in \eqref{eq:estadosX-real} by
\begin{subequations}\label{eq:Xstates-Tij=abcdrs}
\begin{align}
    T_{30}&=a+b-c-d,\\
    T_{03}&=a-b+c-d,\\
    T_{11}&=2(s+r),\\
    T_{22}&=2(s-r),\\
    T_{33}&=a-b-c+d.
\end{align}
\end{subequations}

We can write the projectors involved in the definitions of the classical correlations measures in terms of local $U(2)$ transformations in the standard spherical parametrization, $\mathbbm{U}_I(\theta_I,\phi_I)$ for $I=A,B$. That is, 
\begin{equation}\label{eq:Ui(theta,phi)}
   \mathbbm{U}_I(\theta_I,\phi_I) = \mqty[
   \cos\qty(\frac{\theta_I}{2})            & -\sin\qty(\frac{\theta_I}{2})  \\
   \sin\qty(\frac{\theta_I}{2})\vb{e}^{i\phi_I}   & \cos\qty(\frac{\theta_I}{2}) \vb{e}^{i\phi_I}],
\end{equation}
so that 
\begin{eqnarray}
    \Pi^{I}_{i}(\theta_I,\phi_I)= \mathbbm{U}_I(\theta_I,\phi_I)\dyad{i}\mathbbm{U}_I^\dagger(\theta_I,\phi_I).
\end{eqnarray}
The optimization of the measures for classical correlations, which is a maximization for $\mathcal{C}^s$ \eqref{eq:Cs-Wu} and a minimization for $\mathcal{C}$ \eqref{eq:C(rho)}, leads to the definition of an optimal basis in terms of the angles $\theta_I^{opt}$ and $\phi_I^{opt}$. In Appendix \ref{app:ClassicalCorrelations}, we present the probabilities $p_{ij}$ in terms of the Fano-Bloch parameters and the $\theta_I^{opt}$ and $\phi_I^{opt}$ \cite{LAQC_Xstates-no_sym}.

Next, we can write the projectors to the complementary local bases as
\begin{eqnarray}
    \tilde{\Pi}^{I}_{i}(\theta_I,\phi_I)= \mathbbm{H}(\Phi_I)\Pi^{I}_{i}(\theta_I^{opt},\phi_I^{opt})\mathbbm{H}^\dagger(\Phi_I),
\end{eqnarray}
so we can calculate the respective $p'_{ij}=\Tr\qty(\tilde{\Pi}_i\otimes\tilde{\Pi}_j\,\rho^{AB})$ and perform the corresponding maximization for the quantum correlation measures $Q^s$ \eqref{eq:Qs-Wu} and $\mathcal{L}$ \eqref{eq:LAQC-definition}.

Given the following expressions,
\begin{subequations}\label{eq:Funciones-AlphaBetaGammaDelta}
\begin{align}
    \alpha_i&= 1 + T_{i0} + T_{0i} + T_{ii},\\
    \beta_i&=1 + T_{i0} - T_{0i} - T_{ii},\\
    \gamma_i&= 1 - T_{i0} + T_{0i} - T_{ii},\\
    \delta_i&= 1 - T_{i0} - T_{0i} + T_{ii},\\
\end{align}
\end{subequations}
and $u(x)=(1+x)\log_2(1+x)+(1-x)\log_2(1-x)$, we can define the following function
\begin{equation}\label{eq:Def-gi}
    \begin{aligned}
        g_i &=  \frac{1}{4} \big(\alpha_i\log_2\alpha_i +\beta_i\log_2\beta_i+ \gamma_i\log_2\gamma_i
        \\
        &\qq{ }+\delta_i\log_2\delta_i\big) -\frac{1}{2} \qty\big[u\qty(T_{0i})+u\qty(T_{i0})].
    \end{aligned}
\end{equation}
With the above expressions, we can write the measure of LAQC for 2-qubit X states as
\begin{equation}\label{eq:EstadosX-LAQC}
    \mathcal{L}\qty(\rho_X) = \max_{i=1,2,3}\qty\big{g_i\qty(T_{i0},T_{0i},T_{ii})}.
\end{equation}
If the Fano-Bloch parameters $T_{ij}$ are given in terms of a set of parameters $\Omega=\qty{\Omega_\mu}$, then the maximization in the above expression also depends on $\Omega$. Therefore, depending on the behavior of the $T_{ij}$ in terms of the $\Omega_\mu$, we can have an LAQC measure that is a piecewise-defined function.

One should be aware that the above maximization also requires us to impose that the resulting function is well-behaved. That is, it must be continuous, finite, and null for classical states, as required by the properties of LAQC \cite{LAQC}. We must point out that these requirements are simply a consequence of the simplified way of presenting the quantifier.

Regarding the symmetric measure $Q^s$ introduced by Wu et al. \cite{Wu_ComplementaryBases}, to the best of our knowledge, there is no exact analytical expression for 2-qubit X states in the literature. Given their closely related definitions, we can apply our methodology \cite{LAQC_Xstates-sym, LAQC_Xstates-no_sym} to determine their measures of classical and quantum correlations. Therefore, a direct calculation using the expressions in Eq. \eqref{eq:pij-EstadosX} of Appendix \ref{app:ClassicalCorrelations} lead to
\begin{equation}\label{eq:EstadosX-Cs}
    C^s\qty(\rho_X) = \max_{i=1,2,3}\qty\big{g_i\qty(T_{i0},T_{0i},T_{ii})}.
\end{equation}
The angles $\theta_I^{opt}$ and $\phi_I^{opt}$ involved in the above expression are $\theta_A^{opt} = \theta_B^{opt} = n\pi$ with $n=0,1$ and $\phi_I^{opt}$ undetermined for $C^s\qty(\rho_X) = g_3$, while $\theta_A^{opt} = \theta_B^{opt} = (2n+1)\pi/2$ are related to the other two $g_i$ functions. For $C^s\qty(\rho_X) = g_1$, we require at least one $\phi_I^{opt} = n\pi$ with $n=0,1,2$,  while for  $C^s\qty(\rho_X) = g_2$, requirement is that at least one $\phi_I^{opt} = (2n+1)\pi/2$  with $n=0,1$.

With the above given angles for each of the $g_i$ functions related to the classical correlations measure $C^s$, we can use the expressions presented in Appendix \ref{app:QuantumCorrelations} in the definition for $Q^s$, given in Eq. \eqref{eq:Qs-Wu}. It is straightforward to realize that we can write $Q^s(\rho_X$) in a compact way, analogous to the one we obtained for the LAQC measure \eqref{eq:EstadosX-LAQC}. Therefore, if the function maximizing the above expression is $g_I$, then the symmetric measure of maximal quantum correlations, $Q^s$, is given by the maximum value between the other two functions. That is,
\begin{equation}\label{eq:EstadosX-Qs}
    Q^s\qty(\rho_X) = \max_{i\neq{I}}\qty\big{g_i\qty(T_{i0},T_{0i},T_{ii})}.
\end{equation}
As mentioned beforehand regarding the LAQC measure, the above expression can be a piecewise-defined function if the Fano-Bloch parameters $T_{ij}$ are defined in terms of a set of varying parameters $\Omega=\qty{\Omega_\mu}$.

Finally, the above presented results for the MUB-QC measures points towards a hierarchy between them. By comparing the above expression for $Q^s$ with the measure for LAQC in Eq. \eqref{eq:EstadosX-LAQC}, we can readily realize that, at least for 2-qubit X states, it follows directly that $\mathcal{L}\qty(\rho_X)\geq{}Q^s\qty(\rho_X)$. Moreover, the LAQC measure for X states, given in Eq. \eqref{eq:EstadosX-LAQC}, is numerically equal to the classical correlation measure $C^s$ \eqref{eq:Cs-Wu} defined by Wu et al. Since in their formulation, the classical correlations measure are maximized for local bipartite measurements, one cannot define alternate local bipartite projective measurements that lead to larger $Q^s$ than $C^s$, a limit established by the Holevo bound, as discussed by Wu et al. regarding the components of the symmetric correlation vector \cite{Wu_ComplementaryBases}.


\section{Non-Markovian Quantum Dephasing Channels}\label{sec:QuantumNoise}

Non-Markovian quantum dynamics \cite{Plenio2014-qNon-Markov} is an active field of research in quantum information science and technology. The assumption of Markovian dynamics imposes restrictions on the interactions studied that are not always realistic, especially when one is interested in real-life technological applications rather than theoretical explorations. In this context, analyzing more general situations where non-Markovian dynamics prevail is essential.

Although non-Markovian decoherence is  generally studied via master equations \cite{Schlosshauer, Schlosshauer2019QuantumDeco}, there are processes that one can analyze using extensions of the Kraus operators for the Markovian case. Specifically, several studies focused on non-Markovian phase-flipping and dephasing channels \cite{Daffer2004-QuantumRTN, Kumar2018-NonMarkovianGameTheory, Thapliyal2017-qCrypto-NonMarkovChannel, Naikoo2019-QI_NonMarkovEvol, Mishra2022-QIP-NonMark-Xstates, Sabale2024Dchannels-RTN-OUN, Aiache2025-Dephasing}. In this section, we focus on the two examples of non-Markovian phase-flipping channels that are the subject of this investigation. 

For qubits, the Phase-Flipping channel is directly related to the Pauli $\mathbbm{Z}$ gate, and we can write the Kraus operators in their canonical form as $\vb{K}_0 = \sqrt{1-p\,} \mathbbm{1}$ and $\vb{K}_1 = \sqrt{p\,}\sigma_z$, where $\sigma_z$ is the corresponding Pauli operator, and $p\in[0,1]$ is the channel parameter. One can interpret it as the probability of the phase being flipped or as representing the time evolution, with $p(t) = 1-\vb{e}^{-\lambda{}t}$. In what follows, we rewrite the Kraus operators as
\begin{equation}\label{eq:Kraus-PhaseFlip}
    \vb{K}_0 = \sqrt{\frac{1+\Lambda(t)}{2}\,}\,\mathbbm{1}\qc \vb{K}_1 = \sqrt{\frac{1-\Lambda(t)}{2}\,}\,\sigma_z,
\end{equation}
where $\Lambda(t)$ depends on the particular noise  being studied.

When modeling a common bath for both subsystems, we can write the resulting state as
\begin{equation}\label{eq:Phi(rho)}
    \Phi(\rho)= \sum_{i,j}\vb{K}_{ij}\,\rho\,\vb{K}_{ij}^\dagger,
\end{equation}
where $\vb{K}_{ij}=\vb{K}_{i}\otimes\vb{K}_{j}$. For a general 2-qubit X state \eqref{eq:estadosX-real}, with Bloch parameters $T_{30}$, $T_{03}$, $T_{11}$, $T_{22}$, and $T_{33}$, the phase flipping operation operators only affect the $T_{11}$ and $T_{22}$ parameters, which become
\begin{equation}\label{eq:T11&T22-Lambda(t)}
    T^{(PF)}_{11}=\Lambda^2(t)\,T_{11}\qc T^{(PF)}_{22}=\Lambda^2(t)\,T_{22}.
\end{equation}
Since these parameters are the ones involved in the off-diagonal elements of $\rho_X$ \eqref{eq:estadosX-real}, when $\Lambda(t)=0$, the state becomes diagonal and therefore classical. For Markovian processes, this only occurs when $t\rightarrow\infty$ and one only observes an asymptotic death of MUB-QC.

\subsection{Random Telegraph Noise}

Random Telegraph Noise (RTN), also called Burst Noise, is a type of quasi-discrete noise that occurs in electronic devices, especially in semiconductors and ultrathin gate oxide films \cite{Simoen2011-RTN, Wold2012ClassicRTN, Mulong2015ImpactsOfRTN}. It is characterized by arbitrary and step-like transitions near two or more voltage or current levels. Mathematically, this classical signal noise is modeled by a stochastic Markovian process called a telegraph or dichotomous random process \cite{West2006ModelingComplex-RTN}.

Although this classical noise and its related statistical processes are Markovian, they can be extrapolated to model random step-like transitions in quantum systems, generally a two-level system. For instance, modeling the random transitions in strong laser-atom interactions can be done via an analogous random process leading to a non-Markovian master equation, a process labeled by Eberly and collaborators as phase telegraph noise \cite{Eberly1984_RTN-Laser&Atom}.

In 2004, Daffer et al. \cite{Daffer2004-QuantumRTN} analyzed master equations defined beyond the canonical GKS generator \cite{Gorini1976-GKS_Eq} by introducing a time-dependent integral operator. Its definition includes a well-behaved, continuous kernel function that accounts for the memory effects involved in the physical model. With this memory kernel, the authors were able to introduce colored noise to their model and depart from a Markovian dynamic.

The authors then studied the conditions under which their non-Markovian master equation defines a completely positive map. In doing so, they discussed the applications to non-Markovian phase damping. Building on these results, Kumar and collaborators \cite{Kumar2018-NonMarkovianGameTheory} adopted a quantum walk perspective to define the noise function $\Lambda(t)$ \eqref{eq:Kraus-PhaseFlip} for RTN. Based on a damped harmonic oscillator, they obtained that
\begin{equation}\label{eq:Lambda(t)-RTN}
    \Lambda(t) = \vb{e}^{-\gamma\,t}\qty[\cos\omega\gamma\,t + \frac{\sin\omega\gamma\,t}{\omega}],
\end{equation}
where $\omega = \sqrt{\qty({2a}/{\gamma})^2 -1 \,}$ is the frequency of the harmonic oscillator, $\gamma$ is the fluctuation rate of the RTN, and $a$ is defined within the autocorrelation function for this type of noise. It accounts for the strength of the system-environment coupling.

As Kumar et al. point out \cite{Kumar2018-NonMarkovianGameTheory}, the above-defined noise function has two distinct regimes, depending on whether $a/\gamma$ is greater or less that $1/4$. In particular, the damped oscillations of the non-Markovian behavior correspond to $a/\gamma>1/4$, while the purely damping behavior occurs for $a/\gamma<1/4$. The authors also emphasize that one can verify the non-Markovian behavior regime using the CP-divisibility \cite{Rivas2010Entang-non_Markov, Plenio2014-qNon-Markov} and backflow criteria \cite{Breuer2009Measure-non-Markov}.

\subsection{Modified Ornstein-Uhlenbeck Noise}

The Ornstein-Uhlenbeck (OU) process is a temporally homogeneous Gauss-Markov process introduced to model the velocity of Brownian motion in a viscous medium \cite{Uhlenbeck1930O-OU-original}. Some generalizations, specifically those associated with Lévi processes, have been applied in other areas such as financial modeling \cite{Maller2009-Gen_OrnsteinUhlenbeck}. A modified version of the OU process refers to a non-Markovian aleatory Gaussian process with a well-defined Markovian regime limit. Quantum mechanically, this type of noise can be observed in the spin degree of freedom of an electron interacting with a stochastically fluctuating magnetic field \cite{Zhang2020-OUN_spinB, Turner2022-OU_spinB}.

The time-dependent channel parameter $\Lambda(t)$ for the modified OU (MOU) noise is \cite{Kumar2018-NonMarkovianGameTheory, Sabale2024Dchannels-RTN-OUN}
\begin{equation}\label{eq:Lambda(t)-MOUN}
    \Lambda(t) = \vb{e}^{-\frac{\Gamma}{2}\qty[(t+\frac{1}{\gamma}\qty(\vb{e}^{-\gamma{}t}-1)]},
\end{equation}
where $\gamma^{-1}$ is the finite correlation time with the environment, and $\Gamma$ is the dephasing relaxation time, also referred to as the $T_2$ time.

For the MOU noise function defined above, the regime only becomes Markovian in the limit $\gamma\rightarrow\infty$, where $\Lambda(t)=\exp\qty(-\Gamma\,t/2)$. As long as $\gamma$ is finite, the dynamics are non-Markovian, since the resulting spectrum is Lorentzian and not a delta function, as is the case in the Markovian regime \cite{Kumar2018-NonMarkovianGameTheory}.


\section{MUB-QC measures under non-Markovian dephasing}\label{sec:Results}

As we already pointed out, the non-Markovian Phase Flip channel will only affect the $T_{11}$ and $T_{22}$ Bloch parameters of an X state, as shown in Eq. \eqref{eq:T11&T22-Lambda(t)}. Since these are the parameters involved in the off-diagonal elements of the state \eqref{eq:estadosX-real}, studying the zeros of the corresponding $\Lambda(t)$ functions equals studying when the state becomes classical, i.e. diagonal in a local basis. As was proven in \cite{LAQC}, a state that is diagonal in a local basis has zero LAQC, and given that $Q^s(\rho_X)\leq \mathcal{L}(\rho_X)$, the symmetric measure $Q^s$ is also necessarily zero for a diagonal state.

For the RTN's $\Lambda(t)$ function \eqref{eq:Lambda(t)-RTN}, a simple calculation shows that for $t\geq0$, it becomes zero when
\begin{equation}
    t=n\pi-\,\frac{\arctan\omega}{\omega\gamma}\qc n=0,1,2,\ldots,
\end{equation}
giving rise to the possibility of sudden death and revival of MUB-QC. 

The fact that we can affirm that we will observe null MUB-QC measures for at least those times for which $\Lambda(t)=0$ raises the question of whether other conditions arise that lead to the sudden death of LAQC in this context. To address this issue, we can rely on the necessary and sufficient conditions for zero LAQC established by Mundarain and Ladrón de Guevara \cite{LAQC}. Given a 2-qubit state $\rho^{AB}$, it will have zero LAQC if and only if the corresponding density matrix written in the optimal basis has null anti-diagonal elements, $\tilde{\rho}_{14}=\tilde{\rho}_{23}=0$, and either $\tilde{\rho}_{12}+\tilde{\rho}_{34}$, $\tilde{\rho}_{13}+\tilde{\rho}_{24}$, or both are zero (see Theorem 3 in Ref. \cite{LAQC}).

If we apply the local unitary transformations $\mathbbm{U}_A\otimes\mathbbm{U}_B$ \eqref{eq:Ui(theta,phi)} to the density matrix $\rho_X$ \eqref{eq:estadosX-real}, we have that

\begin{subequations}\label{eq:R14&R23-opt}
    \begin{align}
        \tilde{\rho}_{14} = \,& \frac{\vb{e}^{-i\phi_{+}}}{4}\, \big[ \Lambda^2(t)\qty(T_{11} - T_{22}\cos\theta_A\cos\theta_B)
        \nonumber\\
        & \qq{ }\; +\sin\theta_A\sin\theta_B\big],\label{eq:R14Xopt}
        \\
        \tilde{\rho}_{23} = \,& \frac{\vb{e}^{-i\phi_{-}}}{4}\, \big[ \Lambda^2(t)\qty(T_{11} + T_{22}\cos\theta_A\cos\theta_B)
        \nonumber\\
        & \qq{ }\; +\sin\theta_A\sin\theta_B\big],\label{eq:R23Xopt}
    \end{align}
\end{subequations}
where $\phi_{\pm}=\phi_A\pm\phi_B$. On the other hand, a direct calculation leads to
\begin{subequations}
    \begin{align}
        \tilde{\rho}_{12}+\tilde{\rho}_{34} = \; \frac{\vb{e}^{-i\phi_{B}}}{2}\,\qty\big[\Lambda^2(t)T_{11}\sin\theta_A +\sin\theta_B],\label{eq:R12+R34opt}
        \\
        \tilde{\rho}_{13}+\tilde{\rho}_{24} = \; \frac{\vb{e}^{-i\phi_{A}}}{2}\,\qty\big[\Lambda^2(t)T_{11}\sin\theta_B +\sin\theta_A]\label{eq:R13+R24opt}.
    \end{align}
\end{subequations}

Given these expressions and relations, and taking into account that the angles $\theta_A, \theta_B, \phi_A$, and $\phi_B$ are fixed by the optimization of the classical correlations measure, we can now affirm that, if the initial state has non-zero LAQC, we will only observe the death of LAQC in this context when $\Lambda(t)=0$.

For the MOU noise, the $\Lambda(t)$ function \eqref{eq:Lambda(t)-MOUN} only tends to zero asymptotically. Therefore, we can affirm that MOUN will not induce the sudden death of LAQC for X states that have non-zero LAQC initially.

\begin{figure}[t]
\centering
    \begin{subfigure}[c]{.23\textwidth}
        \includegraphics[width=\textwidth]{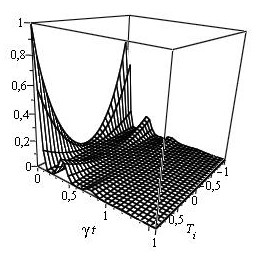}
        \caption{RTN}\label{subfig:g_i-RTN}
    \end{subfigure} 
    \hfill 
    \begin{subfigure}[c]{.23\textwidth}
        \includegraphics[width=\textwidth]{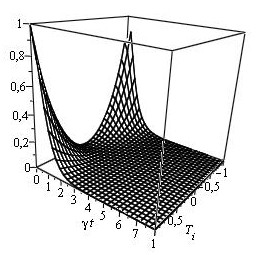}
        \caption {MOUN}\label{subfig:g_i-MOUN}
    \end{subfigure}
\caption{Time evolution of the $g_1$ and $g_2$ functions for a general 2-qubit X state under random telegraph noise (RTN) in units of $\gamma{}t$ and $a = 4\gamma$ (\subref{subfig:g_i-RTN}), and under the modified Ornstein-Uhlenbeck noise (MOUN), in units of $\gamma{t}$ and $\Gamma=\gamma$ (\subref{subfig:g_i-MOUN}).\label{fig:Xgen-RQC-RTN&MOU}}
\end{figure}

In Figure \ref{fig:Xgen-RQC-RTN&MOU}, we show the results for the time evolution of the $g_1$ and $g_2$ functions for a general 2-qubit X state under RTN (\ref{subfig:g_i-RTN}) and MOU noise (\ref{subfig:g_i-MOUN}). As discussed in \cite{Kumar2018-NonMarkovianGameTheory}, RTN induces an oscillatory behavior associated with information backflow, which does not occur in its Markovian analogue. Meanwhile, MOU noise lacks this characteristic and no revival of entanglement or MUB-QC is observed.

To better illustrate these results, we study four different families of 2-qubit X states. Starting with the highly symmetrical Werner states \cite{Werner}, we also analyze the so-called Maximally Nonlocal Mixed States \cite{Batle2011Nonlocality, Fan2019Inequality} (MNLMS) of the Bell-Diagonal subtype, and Maximally Entangled Mixed States  \cite{Munro2001Maximizing} (MEMS). Finally, we study a mixed symmetric separable state \cite{LAQC_Swap} with non-zero LAQC measure.


\subsection{Werner states}

Introduced in 1989, Werner states are defined as the family of bipartite states that are invariant under local unitary transformations. For 2-qubit Werner states, they are usually written as
\begin{equation}\label{eq:rho-Werner}
    \rho_w= z\dyad{\psi^-}+\frac{1-z}{4}\,\mathbbm{1}_4,
\end{equation}
where $\ket{\psi^-}=\frac{1}{\sqrt{2\,}}\qty(\ket{01}-\ket{10})$ is the singlet state, one of the four maximally entangled Bell states, and $z\in[0,1]$. For such states, their LAQC measure is given by
\begin{equation}\label{eq:Werner-RQC}
    \mathcal{L}(\rho_w)= \frac{1+z}{2}\log_2(1+z) + \frac{1-z}{2}\log_2(1-z),
\end{equation}
and it is straightforward to realize from Eq. \eqref{eq:EstadosX-Qs} that $Q^s\qty(\rho_w)= \mathcal{L}(\rho_w)$.

\begin{figure}[b]
\centering
\includegraphics[width=.35\textwidth]{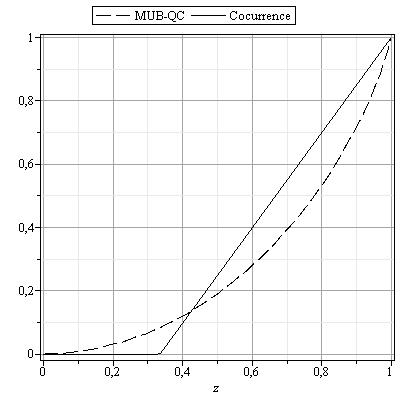}
\caption{Concurrence (solid) and MUB-defined Quantum Correlations (dashed) for Werner states.\label{fig:Werner-RQC&Conc}}
\end{figure}

It is well known that Werner states are separable for $z\leq1/3$. Using \eqref{eq:Concurrencia_X}, it is straightforward to verify that
\begin{equation}\label{eq:Werner-Conc}
    C(\rho_w)=\frac{1}{2}\,\max\qty{0,3z-1}.
\end{equation}
In Figure \ref{fig:Werner-RQC&Conc}, we present the graphs for concurrence (continuous line) and MUB-QC (dashed line) for Werner states. We can observe that for $z\sim 0.421499471$, its entanglement and the two MUB-QC measures are numerically equal.

\begin{figure}[ht]
\centering
\begin{subfigure}[c]{.23\textwidth}
    \centering
    \includegraphics[width=\textwidth]{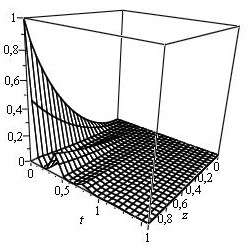}
    \caption{$\mathcal{L}[\rho_w(t)]$ and $Q^{s}[\rho_w(t)]$}
    \label{subfig:Werner-RTN-LAQC}
\end{subfigure}
 \hfill
 \begin{subfigure}[c]{.23\textwidth}
    \centering
    \includegraphics[width=\textwidth]{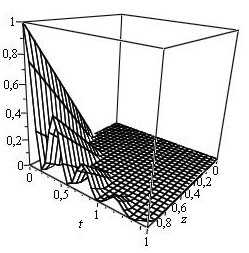}
    \caption{$\mathcal{C}[\rho_w(t)]$}
    \label{subfig:Werner-RTN-Conc}
 \end{subfigure}
\caption{Time evolution of  MUB-QC (\subref{subfig:Werner-RTN-LAQC}) and Concurrence (\subref{subfig:Werner-RTN-Conc}) for Werner states under random telegraph noise (RTN) in units of $\gamma{}t$ with $a = 4\gamma$.\label{fig:Werner-RQC&Conc-RTN}}
\end{figure}

The temporal evolution of these quantum correlations for a Werner state under random telegraph noise (RTN) is shown in Figure \ref{fig:Werner-RQC&Conc-RTN}, for $a = 4\gamma$, that is, $\omega=3\sqrt{7}$ \eqref{eq:Lambda(t)-RTN}. We can readily observe that the damping of MUB-QC (Figure \ref{subfig:Werner-RTN-LAQC}) occurs rapidly, with revival peaks fading sooner than they do for entanglement (Figure \ref{subfig:Werner-RTN-Conc}) for $z$ large enough. Nevertheless, since
\begin{equation}\label{eq:Werner-Conc-RTN}
    \mathcal{C}\qty(\rho_w^{(RTN)}) = \frac{1}{2}\,\max\qty{0, \qty\big[1+2\Lambda^2(t)]z-1},
\end{equation}
entanglement revival peaks will fall below those of these MUB-QC as $z$ decreases.

\begin{figure}[b]
\centering
    \begin{subfigure}[c]{.23\textwidth}
        \centering
        \includegraphics[width=\textwidth]{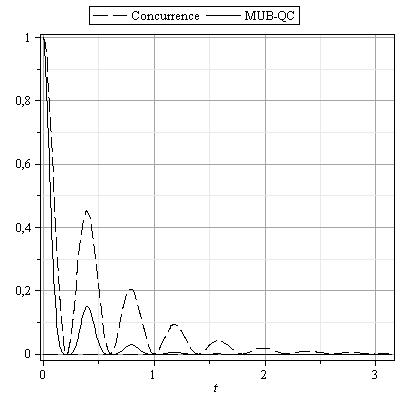}
        \caption{$\rho(t=0)=\dyad{\psi^-}$}
        \label{subfig:PsiMenos-RQC_Conc-RTN}
    \end{subfigure}
    \hfill 
    \begin{subfigure}[c]{.23\textwidth}
        \centering
        \includegraphics[width=\textwidth]{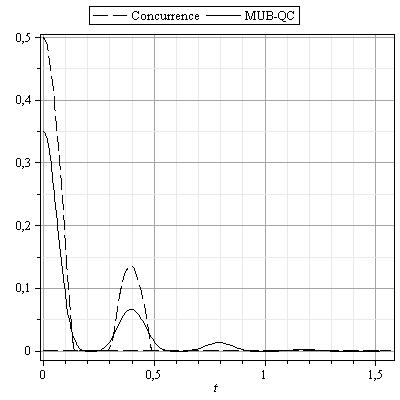}
         \caption{$\rho(t=0)=\rho_w(z=2/3)$}
        \label{subfig:Werner(z=2/3)-RQC&Conc-RTN}
    \end{subfigure}
\caption{Comparison of the time evolution of Concurrence (dashed) and MUB-QC (solid) under rnadom tepegraph noise (RTN) for $\ket{\psi^-}$ (\subref{subfig:PsiMenos-RQC_Conc-RTN}) and a Werner state with $z=2/3$ (\subref{subfig:Werner(z=2/3)-RQC&Conc-RTN}) in units of $\gamma{}t$ and $a = 4\gamma$.}
\label{fig:PsiSinglet-RQC&Conc-RTN}
\end{figure}

When comparing the behavior of the singlet state, $\ket{\psi^-}$, shown in Figure \ref{subfig:PsiMenos-RQC_Conc-RTN}, the difference in peaks is more evident. Entanglement revivals achieve higher peaks in the initial time range $0\leq\gamma{t}\leq2$, while MUB-QC revivals are almost undetectable for $\gamma{t}>1$. On the other hand, Figure \ref{subfig:Werner(z=2/3)-RQC&Conc-RTN} presents the time evolution of a state with a Werner parameter $z=2/3$. As we mentioned earlier, the entanglement measure is larger than that for the two MUB-QC analyzed, but this behavior rapidly flips. While the first revival peak for concurrence is larger, there are no further entanglement revivals for this state, although two more revivals are detectable for the MUB-QC measures at this scale.

The behavior observed for Werner states and discussed above can be explained by looking at the quantitative relation between concurrence and the MUB-QC measure. Given a value of the parameter $z$, we can consider a vicinity around that point in the MUB-QC measure curve presented in Figure \ref{fig:Werner-RQC&Conc}. Near $z=1$, the curve departs from the concurrence line up until $z=7/9$, where the behavior flips and starts to get closer. As the curve departs from the values of concurrence, the revival oscillations for MUB-QC are dominated by the tendency of departure. Once the system reaches the threshold of $z=7/9$, the behavior switches and one will eventually observe larger revivals for the MUB-QC measure than for concurrence, even if the first revival still follows the original behavior, as is the case for $\rho_w(z=2/3)$ in Figure \ref{subfig:Werner(z=2/3)-RQC&Conc-RTN}.

\begin{figure}
\centering
\begin{subfigure}[t]{.5\textwidth}
    \begin{subsubfigure}[c]{.48\textwidth}
    \includegraphics[width=\textwidth]{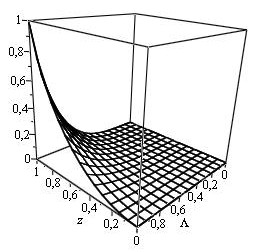}
    \caption{$\mathcal{L}[\rho_w(t)]$ and $Q^{s}[\rho_w(t)]$}\label{subsubfig:Werner-Deph-QC}
    \end{subsubfigure}
     \hfill 
     \begin{subsubfigure}[c]{.48\textwidth}
         \includegraphics[width=\textwidth]{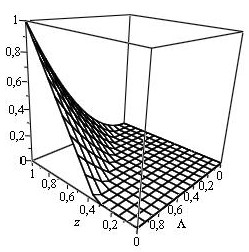}
         \caption{$\mathcal{C}[\rho_w(t)]$}\label{subsubfig:Werner-Deph-Conc}
     \end{subsubfigure}
    \caption{Markovian dephasing}\label{subfig:MarkovDeph}
\end{subfigure}

\begin{subfigure}[t]{.5\textwidth}
\setcounter{subsubfigure}{0}
    \begin{subsubfigure}[c]{.48\textwidth}
    \includegraphics[width=\textwidth]{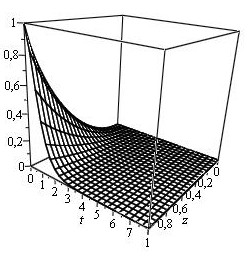} 
    \caption{$\mathcal{L}[\rho_w(t)]$ and $Q^{s}[\rho_w(t)]$}\label{subsubfig:Werner-MOUN-QC}
    \end{subsubfigure}
    \hfill
    \begin{subsubfigure}[c]{.48\textwidth}
        \includegraphics[width=\textwidth]{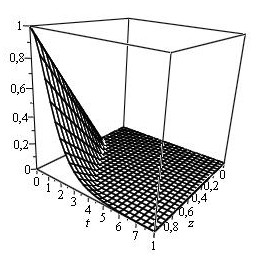}
        \caption{$\mathcal{C}[\rho_w(t)]$}\label{subsubfig:Werner-MOUN-Conc}
    \end{subsubfigure}
    \caption{MOU noise dephasing}\label{subfig:MOUNDeph}
\end{subfigure}
\caption{Time evolution of LAQC and $Q^{s}$ (\subref{subsubfig:Werner-Deph-QC} and \subref{subsubfig:Werner-MOUN-QC}) and Concurrence (\subref{subsubfig:Werner-Deph-Conc} and \subref{subsubfig:Werner-MOUN-Conc}) for Werner states under the Markovian phase-flip channel (\subref{subfig:MarkovDeph}) and the modified Ornstein–Uhlenbeck noise (\subref{subfig:MOUNDeph}) in units of $\gamma{}t$ and $\Gamma = \gamma$.\label{fig:Werner-RQC&Conc-MOUN&PhF}}
\end{figure}

On the other hand, for the modified Ornstein–Uhlenbeck noise (MOUN), the qualitative behavior of both quantum correlation measures is not significantly different from the Markovian case. In  Figure \ref{fig:Werner-RQC&Conc-MOUN&PhF}, we present the Markovian Phase-Flipping channel for the MUB-QC (\ref{subsubfig:Werner-Deph-QC}) and entanglement \eqref{subsubfig:Werner-Deph-Conc} measures in the top row \eqref{subfig:MarkovDeph}. In the bottom row \eqref{subfig:MOUNDeph} of said Figure, we present the temporal evolution under MOUN, with $\Gamma =\gamma$, in terms of $\gamma{}t$, with \eqref{subsubfig:Werner-MOUN-QC} corresponding to the MUB-QC measure and \eqref{subsubfig:Werner-MOUN-Conc}, to concurrence. 

\subsection{Maximally Nonlocal and Maximally Entangled Mixed States}

Introduced in 2011 by Batle and Casas \cite{Batle2011Nonlocality}, maximally nonlocal mixed states (MNMS) are defined as those that maximize the violation of a Bell inequality \cite{Bell1964-OnEPRparadox} for a given purity ($\pi(\rho)=\Tr\rho^2 \leq1$). For instance, when considering the Bell-CHSH inequality \cite{Clauser1969-CHSH_Ineq}, we have
\begin{equation}\label{eq:rho-MNMS}
    \rho_{MNMS} = \dyad{\phi^{+}} -\frac{1-x}{2} \qty(\dyad{00}{11}+\dyad{11}{00})
\end{equation}
with $\ket{\phi^{+}}=\frac{1}{\sqrt{2}}\qty(\ket{00}+\ket{11})$ and $0<x\leq1$ as an MNMS \cite{Mishra2022-QIP-NonMark-Xstates, Nunavath2023-OpenSysDyn-X}. It should be noted that the above defined states belong to the Bell Diagonal type. That is, they have null local Bloch parameters and their reduced operators are therefore maximally mixed.

On the other hand, maximally entangled mixed states were introduced by Munro et al. \cite{Munro2001Maximizing} in 2001 and are defined as 2-qubit mixed states that, for a given linear entropy, have the maximal concurrence possible. Their density matrix is given by
\begin{equation}\label{eq:rho-MEMS}
    \rho_{MEMS} =\frac{1}{2} \mqty(\,2\chi &0 &0 &x\\
                        0& 2-4\chi &0 &0\\
                        0 &0 &0 &0\\
                        x&0 &0 &2\chi\,),
\end{equation}
where $\chi=\chi\qty(x)$ is a function of the parameter $x$ defined as
\begin{equation}\label{eq:Gamma-rho_MEMS}
    \chi = \left\{
    \begin{aligned}
       \; &\frac{1}{3}\qc 0\leq{x}<2/3.\\
        &\frac{x}{2}\qc 2/3\leq{x}\leq1.
    \end{aligned}
    \right.
\end{equation}
It should be noted that, contrary to the previous cases, their reduced states are not maximally mixed. Their local Bloch parameters are non-zero for all values of $x$ as they are given by $T_{30}=-T_{03}=1-2\chi(x)$ and are therefore antisymmetric X states.

\begin{figure}[b]
\centering
\includegraphics[width=.35\textwidth]{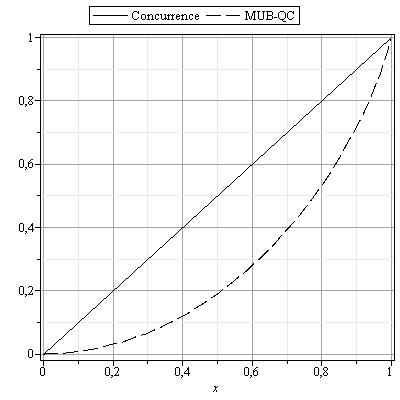}
\caption{Concurrence (solid) and LAQC and $Q^{s}$ (dashed) for maximally nonlocal and maximally entangled mixed states.\label{fig:MNLMS-RQC&Conc}}
\end{figure}

For both the above-defined mixed states, their LAQC measure is given by
\begin{equation}\label{eq:MNLMS-RQC}
    \mathcal{L}(\rho_{MS})= \frac{1+x}{2}\log_2(1+x) + \frac{1-x}{2}\log_2(1-x),
\end{equation}
analogous to the previous result, and $Q^s\qty(\rho_{MS})= \mathcal{L}(\rho_{MS})$. For the rest of this section, we use $\rho_{MS}$ to denote either $\rho_{MNMS}$ or $\rho_{MEMS}$ in results that apply to both states.

In Figure \ref{fig:MNLMS-RQC&Conc}, we present the graphs for concurrence (continuous line) and MUB-QC measures (dashed line) for the above-defined MNMS and MEMS. Contrary to Werner states, these are only separable for $x=0$; their concurrence is linear in the state's parameter, and $\mathcal{C}(\rho_{MS})\geq \mathcal{L}(\rho_{MS})$, where the equality holds only for the separable state, which also happens to be classical.

\begin{figure}[t]
\centering
    \begin{subfigure}[c]{.23\textwidth}
        \includegraphics[width=\textwidth]{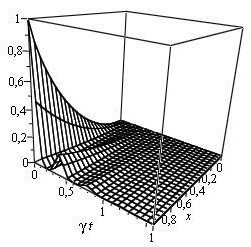}
        \caption{$\mathcal{L}[\rho_{MS}(t)]$ and $Q^{s}[\rho_{MS}(t)]$}\label{subfig:MNMS-QC-MUB-RTN}
    \end{subfigure}
\hfill 
    \begin{subfigure}[c]{.23\textwidth}
        \includegraphics[width=\textwidth]{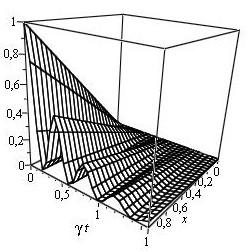}
        \caption{$\mathcal{C}[\rho_{MS}(t)]$}\label{subfig:MNMS-Conc-RTN}
    \end{subfigure}
\caption{Time evolution under random telegraph noise (RTN) of LAQC and $Q^{s}$ (\subref{subfig:MNMS-QC-MUB-RTN}) and Concurrence (\subref{subfig:MNMS-Conc-RTN}) for maximally nonlocal and maximally entangled mixed states in units of $\gamma{}t$ and $a = 4\gamma$.\label{fig:MNLMS-RQC&Conc-RTN}}
\end{figure}

The temporal evolution of concurrence \eqref{subfig:MNMS-Conc-RTN} and the two MUB-QC measures \eqref{subfig:MNMS-QC-MUB-RTN} analyzed for these states under random telegraph noise (RTN) is shown in Figure \ref{fig:MNLMS-RQC&Conc-RTN}, for $a = 4\gamma$, that is, $\omega=3\sqrt{7}$ \eqref{eq:Lambda(t)-RTN}. 

\begin{figure}[b]
    \centering
    \begin{subfigure}[c]{0.23\textwidth}
        \includegraphics[width=\textwidth]{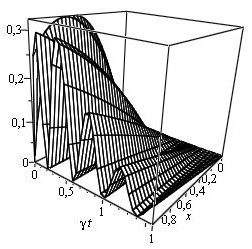}
        \caption{$\mathcal{C}[\rho_{MS}(t)] - \mathcal{Q}^s[\rho_{MS}(t)]$}\label{subfig:MS_C-QS}
    \end{subfigure}
 \hfill 
    \begin{subfigure}[c]{0.23\textwidth}
        \includegraphics[width=\textwidth]{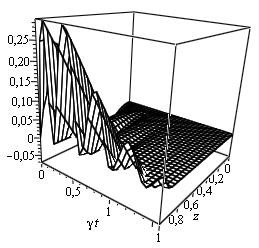}
        \caption{$\mathcal{C}[\rho_w(t)] - \mathcal{Q}^s[\rho_w(t)]$}\label{subfig:Werner_C-QS}
    \end{subfigure}
    \caption{Surface depicting the difference $\mathcal{C}(\rho) - \mathcal{Q}^s(\rho)$ for maximally nonlocal mixed states (\subref{subfig:MS_C-QS}) and Werner states (\subref{subfig:Werner_C-QS}) evolving under random telegraph noise (RTN) in units of $\gamma{}t$ and $a = 4\gamma$.\label{fig:Werner_MNMS-Diff_Conc&RQC}}
\end{figure}

As with the previous case, we can readily observe that the damping occurs more rapidly for the MUB-QC measures than for concurrence. Nevertheless, we now have that the surface representing the time evolution of concurrence is above that of $Q^{s}$ and the LAQC measure for all times, as can be seen in Figure \ref{fig:Werner_MNMS-Diff_Conc&RQC}, where we present surfaces depicting the difference between concurrence and the MUB-QC measure for MS \eqref{subfig:MS_C-QS}, and Werner states \eqref{subfig:Werner_C-QS}. 

From the above-mentioned surfaces, it is clear that for MNMS and MEMS, states that have a higher entanglement than the $Q^{s}$ and LAQC for all values of their defining parameters, this relation is held throughout the decoherence-inducing interaction with RTN. On the other hand, this is not the case for Werner states. The surface in Figure \ref{subfig:Werner_C-QS} has a region for $z<0.421$ where the loss of entanglement is greater than that of the two analyzed MUB-QC measures for all times. For $0.421<z<0.575$, once the loss of entanglement is greater than that of $Q^{s}$ and LAQC, the surface in Figure \ref{subfig:Werner_C-QS} remains negative. For values above $z=0.575$, the behavior is oscillatory with the positive sectors being dominant. Nevertheless, some regions remain negative up until near $z=0.9$.


\subsection{Mixed Symmetric Separable State}

While studying the persistence of LAQC in a standard quantum correlation swapping protocol \cite{LAQC_Swap}, Albrecht introduced a mixed symmetric 2-qubit state that is separable for all values of its defining parameter,
\begin{equation}\label{eq:rho-Xsep}
    \rho_{s} = \frac{1}{4}\,\mqty(s &0 &0 &s\\
                        0& s &s &0\\
                        0 &s &s &0\\
                        s&0 &0 &4-3s),
\end{equation}
where $0\leq{}s\leq1$. One can easily verify that the concurrence for this state is zero for all values of $s$ and that for $s=0$, the state becomes $\ket{11}$. Contrary to the previous cases presented, the LAQC measure and $Q_s$ of $\rho_s$ are not equal and involve the functions $g_1(T_{11})$ and $g_3(T_{30},T_{03},T_{33})$, with a crossover at $s\sim 0.3520$. The non-zero parameters of the Fano form \eqref{eq:2-qubitFanoForm} are 
\begin{equation}\label{eq:rho_s-Bloch}
        T_{30}=T_{03}=s-1\qc T_{11}=s\qc T_{33}=1-s,
\end{equation}
and we can write the corresponding MUB-QC measures of this state as
\begin{equation}
    \mathcal{L}(\rho_s) =\max(g_1,g_3)\qc Q^{s}(\rho_s) =\min(g_1,g_3),
\end{equation}
where
\begin{subequations}
\begin{align}
    g_1 = \, & \frac{1+s}{2}\log_2\qty(1+s)+ \frac{1-s}{2}\log_2\qty(1-s), \\
    g_3 =\, & \frac{4-3s}{4}\log_2(4-3s) -\frac{s}{4}\log_2s\nonumber\\
    &\qq{ } -\qty(2-s)\log_2\qty(2-s).
\end{align}
\end{subequations}

\begin{figure}
    \centering
    \begin{subfigure}[c]{0.23\textwidth}
        \includegraphics[width=\textwidth]{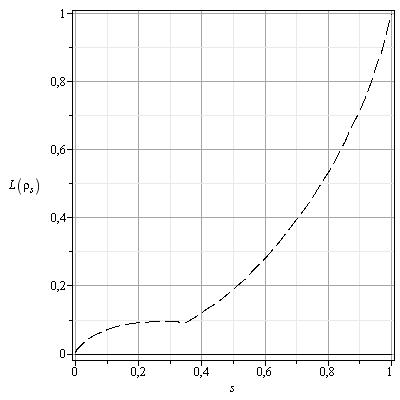}
        \caption{$\mathcal{L}\qty(\rho_{s})$}\label{subfig:rhoS-LAQC}
    \end{subfigure}
    \hfill 
    \begin{subfigure}[c]{0.23\textwidth}
        \includegraphics[width=\textwidth]{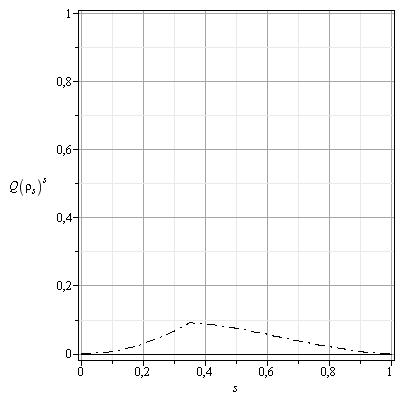}
        \caption{$Q^{s}\qty(\rho_{s})$}\label{subfig:rhoS-Qs}
    \end{subfigure}
    \caption{MUB-QC measures for the mixed symmetric separable state $\rho_s$: LAQC (\subref{subfig:rhoS-LAQC}) and $Q^{s}$ (\subref{subfig:rhoS-Qs}).\label{fig:rho_S-LAQC&Qs}}
\end{figure}

\begin{figure}[b]
    \centering
    \begin{subfigure}[c]{.23\textwidth}
        \includegraphics[width=\textwidth]{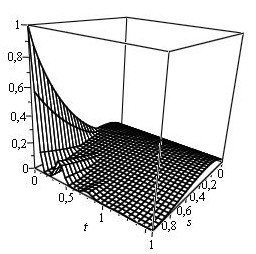}
        \caption{$\mathcal{L}[\rho_{s}(t)]$}\label{subfig:rhoS-LAQC-RTN}
    \end{subfigure}
    \hfill
    \begin{subfigure}[c]{.23\textwidth}
        \includegraphics[width=\textwidth]{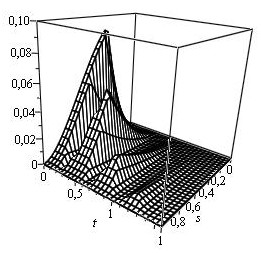}
        \caption{$Q^{s}[\rho_{s}(t)]$}\label{subfig:rhoS-Qs-RTN}
    \end{subfigure}
    \caption{Time evolution of the MUB-QC measures for the mixed symmetric separable state $\rho_s$ subject to a random telegraph noise (RTN) dephasing channel: LAQC (\subref{subfig:rhoS-LAQC-RTN}) and $Q^{s}$ (\subref{subfig:rhoS-Qs-RTN}).\label{fig:rho_S-RTN-LAQC&Qs}}
\end{figure}

In Figure \ref{fig:rho_S-LAQC&Qs}, we present the graphs for LAQC
(dashed line, \ref{subfig:rhoS-LAQC}) and the symmetric measure $Q^{s}$
(dash-dotted line, \ref{subfig:rhoS-Qs}) for this separable mixed state. Regarding the surfaces depicting the time evolution under RTN of these MUB-QC measures, we present them in Figure \ref{fig:rho_S-RTN-LAQC&Qs}, with the left graph corresponding to the LAQC measure (\ref{subfig:rhoS-LAQC-RTN}) and the right  one to $Q^s$ (\ref{subfig:rhoS-Qs-RTN}). It should be noted that the scale for $Q^s$ in Figure \ref{subfig:rhoS-Qs-RTN} was adjusted to better visualize the effects of RTN on this measure. On the other hand, the state $\rho_s^{(RTN)}$ remains separable for all times, as expected.

\subsection{General Remarks}

In \cite{LAQC}, Mundarain and Ladrón de Guevara proved that the LAQC of the original state correspond to the correlations of local measurements, performed in the local maximally entangled system-ancilla basis, when the entanglement activated by Piani's protocol \cite{Piani_2011-EntActiv} is minimal. The results for Werner states, particularly those shown in Figure \ref{subfig:Werner(z=2/3)-RQC&Conc-RTN}, serve as an example of state dynamics for which the LAQC measure reveals the existence of an activatable quantum resource, even when the states are no longer entangled. Moreover, there are separable states with non-zero LAQC, such as the family of mixed symmetric separable states $\rho_s$ defined in Eq. \eqref{eq:rho-Xsep}, that exhibit non-trivial behavior under RTN dephasing.  

All of the above provides Local-Available Quantum Corrrelations (LAQC) with a clear physical meaning as an indicator of the robustness of local measurement correlations. Its close relationship to Piani's protocol, together with the hierarchy between the latter and quantum discord, as shown by Mundarain and Ladrón de Guevara \cite{LAQC}, suggests that a non-zero LAQC can also be turned into a usable quantum resource such as distillable entanglement. 

\section{Summary and Conclusions}\label{sec:Conclusions}

We studied the dynamics of two measures of quantum correlations defined in terms of mutually unbiased bases (MUB-QC)  for bipartite qubit systems under a non-Markovian quantum dephasing channel. As channel functions, we used those that model random telegraph noise (RTN) and modified Ornstein-Uhlenbeck noise (MOUN). We focused our analysis on X states, which are of special interest due to their theoretical and practical relevance.

Using our previous approach to determine an analytical exact expression for $\mathcal{L}(\rho_X)$, which quantifies local-available quantum correlations (LAQC), a type of MUB-QC, we derived an exact analytical result for the symmetric measure $\mathcal{Q}^s$ introduced by Wu et al. \cite{Wu_ComplementaryBases} in 2-qubit X states.  In doing so, we also showed that $\mathcal{L}(\rho_X)\geq\mathcal{Q}^s(\rho_X)$ for 2-qubit X states.

Since the Kraus operators involved in this quantum channel only affect the off-diagonal elements of the initial state, we were able to draw general conclusions for the dynamics of the two types of MUB-QC measures analyzed under RTN and MOUN. In particular, we observe the sudden death of LAQC driven by RTN, a result we had not seen in our previous analysis using Markovian quantum channels. Moreover, RTN quantum dephasing also led to the revival of the two MUB-QC studied in this paper.

We analyzed four physically relevant families of X states to illustrate our general results: Werner states, maximally nonlocal (MNMS) and maximally entangled mixed states (MEMS), and a separable symmetric X state. To compare the behavior of the MUB-QC measures, we chose concurrence as a \textit{bona fide} measure of entanglement.
 
While there is a range of Werner states for which the MUB-QC measures lie above concurrence, that is not the case for MNMS and MEMS. This was also reflected in how the deaths and revivals of both measures behave for these states. In Werner states, we found that the revivals of the MUB-QC measures are greater than those of concurrence for a wide range of the parameter $z$. Even more so, for most values of $z$ for which the concurrence is greater than the MUB-QC measures, after the behavior switched and the loss of concurrence is greater, for up to $z\sim0.9$, we have a larger measure of MUB-QC than that of entanglement. For MNMS and MEMS, concurrence is larger than the measures of MUB-QC for all $x$ and throughout the system's evolution.

Finally, we analyzed a symmetric X state that is separable for all values of its parameter $s$ but has non-zero quantum correlations. Contrary to all previous states analyzed, $Q^s(\rho_s)$ and $\mathcal{L}(\rho_s)$ exhibit distinct behaviors and are only equal for two values of $s$. The two MUB-QC measures studied exhibit death and revival throughout their evolution, as was observed in previous cases, while remaining separable at all times.


\section*{Acknowledgments}

The authors would like to acknowledge the support of their research groups at Universidad Simón Bolívar, Venezuela. Albrecht’s group is GID-30, \emph{Teoría de Campos y Óptica Cuántica}, and Bellorín’s group is GID-35, \emph{Materia Condensada y Sistemas Complejos}.


\section*{Declarations}

No funding was received to assist with the preparation of this manuscript. 

No GenAI tools were used in the conception and design of this study or to prepare the manuscript, complete the reference list, or generate the figures. A free web-based AI-assisted proofreader was used as a tool to assist the authors in improving grammar, spelling, and wording in the original draft. The authors take full responsibility for the final content and attest that it represents their own intellectual effort.

\section*{Data availability statement}

The authors declare that no datasets were generated or analyzed for the preparation of this article. Therefore, no underlying data are available.

\appendix
\section{Determining the Classical Correlation Measures}\label{app:ClassicalCorrelations}

By applying the local transformations $\mathbbm{U}_I(\theta_I,\phi_I)$ \eqref{eq:Ui(theta,phi)} to $\rho_X$ \eqref{eq:estadosX-real}  in the Fano form \eqref{eq:2-qubitFanoForm},  we obtain, after some algebraic manipulation, the following expressions for $p_{ij}$ \cite{LAQC_Xstates-no_sym}:
\begin{subequations}\label{eq:pij-EstadosX}
\begin{align}
     p_{00}=\,& \frac{1}{4}\Big[1+T_{30}\cos\theta_A +T_{03}\cos\theta_B +T_{33}\cos\theta_A\cos\theta_B\nonumber
            \\ &\qq{}+\sin\theta_A\sin\theta_B\big(T_{11}\cos\phi_A\cos\phi_B\nonumber
        \\ 
        &\qq{\hspace{10mm}}+T_{22}\sin\phi_A\sin\phi_B\big)\Big],
\\
    p_{01}=\,& \frac{1}{4}\Big[1+T_{30}\cos\theta_A -T_{03}\cos\theta_B -T_{33}\cos\theta_A\cos\theta_B \nonumber
        \\ 
        &\qq{}-\sin\theta_A\sin\theta_B\big(T_{11}\cos\phi_A\cos\phi_B\nonumber
        \\ 
        &\qq{\hspace{10mm}}+T_{22}\sin\phi_A\sin\phi_B\big)\Big],
\\
    p_{10}=\,& \frac{1}{4}\Big[1-T_{30}\cos\theta_A +T_{03}\cos\theta_B -T_{33}\cos\theta_A\cos\theta_B \nonumber
        \\ 
        &\qq{}-\sin\theta_A\sin\theta_B\big(T_{11}\cos\phi_A\cos\phi_B\nonumber
        \\ 
        &\qq{\hspace{10mm}} +T_{22}\sin\phi_A\sin\phi_B\big)\Big],
\\
    p_{11}
        =\,& \frac{1}{4}\Big[1-T_{30}\cos\theta_A -T_{03}\cos\theta_B+T_{33}\cos\theta_A\cos\theta_B\nonumber
        \\ 
        &\qq{}+\sin\theta_A\sin\theta_B\big(T_{11}\cos\phi_A\cos\phi_B\nonumber
        \\ 
        &\qq{\hspace{10mm}}+T_{22}\sin\phi_A\sin\phi_B\big)\Big].
\end{align}
\end{subequations}

For the reduced states $\rho_I$, we have that $p_i^{I}$ are given by \cite{LAQC_Xstates-no_sym}
\begin{subequations}
\begin{align}
    p_{\smqty{0\\1}}^{(\vb{A})}&=\frac{1}{2}\qty(1\pm{}T_{30}\cos\theta_A),
\\
    p_{\smqty{0\\1}}^{(\vb{B})}&=\frac{1}{2}\qty(1\pm{}T_{03}\cos\theta_B).
\end{align}
\end{subequations}

These are the $p_{ij}$ involved in the definitions of the classical correlations measure for $Q^s$ \eqref{eq:Cs-Wu} and LAQC \eqref{eq:C(rho)}. Therefore, we can verify that $C^s(\rho_X)$ is as given by Eq. \eqref{eq:EstadosX-Cs}, while for LAQC we have
\begin{equation}\label{eq:LAQC-C(rhoX)}
    C\qty(\rho_X) = \left\{
    \begin{aligned}
        0\qc & \qty|T_{30}|\neq\qty|T_{03}|
        \\
        \min_{i=1,2,3}\qty\big[g_i]\qc & \qty|T_{30}|=\qty|T_{03}|
    \end{aligned}
    \right. .
\end{equation}

The optimization involved in these measures defines the optimal computational basis by fixing the parametrization angles $\theta_I^{(opt)}$ and $\phi_I^{(opt)}$, although the latter can, in general, be undetermined until one maximizes the quantum correlation measure.

\section{Determining the MUB-QC Measures}\label{app:QuantumCorrelations}

To determine $Q^s$ and the LAQC measure, one has to calculate the $p'_{ij}$ required in Eq. \eqref{eq:Qs-Wu} and \eqref{eq:LAQC-definition}, respectively. These probabilities are given in terms of the angles defining the optimal basis, $\theta_I^{(opt)}$ and $\phi_I^{(opt)}$ with $I=A,B$. Depending on the MUB-QC measure one is interested in calculating, these angles are defined by maximizing or minimizing the $p_{ij}$'s presented in Appendix \ref{app:ClassicalCorrelations}.

Since the general expressions $p'_{ij}\qty(\Phi_I, \theta_I^{(opt)}, \phi_I^{(opt)})$ are lengthy and tedious to read, we present the expressions for given selection of the optimal basis parametrization angles $\theta_I^{(opt)}$ and $\phi_I^{(opt)}$. We start by defining 
\begin{equation}\label{eq:Psi-pm}
    \Psi_I^\pm=\Phi_I \pm \phi_I\qc I=A,B,
\end{equation}
since for various optimal bases, the $\phi_I$ remain undetermined in the first optimization.

When the optimal basis is the original basis, that is, for $\theta_A^{(opt)}=\theta_B^{(opt)}= 2n\pi$ ($n=0,1)$ and $\phi_I^{(opt)}$ undetermined, we have
\begin{subequations}\label{eq:p'_ij-theta0}
    \begin{align}
        p_{00}' =\; & \frac{1}{4} \big( 1 +T_{11}\cos\Psi_A^+ \cos\Psi_B^-\notag\\
        &\qq{ } +T_{22}\sin\Psi_A^+\sin\Psi_B^-\big) =p_{11}',
        \\
        p_{01}' =\; & \frac{1}{4} \big( 1 -T_{11}\cos\Psi_A^+ \cos\Psi_B^-\notag\\
        &\qq{ } -T_{22}\sin\Psi_A^+\sin\Psi_B^-\big)=p_{10}'.
    \end{align}
\end{subequations}
For $\theta_A^{(opt)}=\theta_B^{(opt)}= \pi$, we have the same expressions, with the exchange $p'_{00} \leftrightarrow p'_{01}$ being the only difference.

When the optimal basis is given by $\theta_A^{(opt)}=n\pi$ and $\theta_B^{(opt)}= (2n+1)\pi$, with $n=0,1)$, and $\phi_I^{(opt)}$ undetermined, we have
\begin{subequations}\label{eq:p'_ij-theta0Pi/2}
    \begin{align}
        p_{00}' =\; & \frac{1}{4} \big( 1 - T_{03}\cos\Phi_B +T_{11}\cos\Psi_A^+ \sin\Phi_B\sin\phi_B\notag
        \\
        &\qq{ } +T_{22}\sin\Psi_A^+ \sin\Phi_B\cos\phi_B\big) =p_{11}',
        \\
        p_{01}' =\; & \frac{1}{4} \big( 1 + T_{03}\cos\Phi_B -T_{11}\cos\Psi_A^+ \sin\Phi_B\sin\phi_B\notag
        \\
        &\qq{ } -T_{22}\sin\Psi_A^+ \sin\Phi_B\cos\phi_B\big) =p_{10}'.
    \end{align}
\end{subequations}
When the optimal basis is given by $\theta_A^{(opt)}=(2n+1)\pi/2$ and $\theta_B^{(opt)}= n\pi$, with $n=0,1$, and $\phi_I^{(opt)}$ undetermined, we have analogous expressions to those above. That is,
\begin{subequations}\label{eq:p'_ij-thetaPi/20}
    \begin{align}
        p_{00}' =\; & \frac{1}{4} \big( 1 - T_{30}\cos\Phi_A +T_{11}\cos\Psi_B^+ \sin\Phi_A\sin\phi_A\notag
        \\
        &\qq{ } +T_{22}\sin\Psi_B^+ \sin\Phi_A\cos\phi_A\big) =p_{11}',
        \\
        p_{01}' =\; & \frac{1}{4} \big( 1 + T_{30}\cos\Phi_A -T_{11}\cos\Psi_B^+ \sin\Phi_A\sin\phi_A\notag
        \\
        &\qq{ } -T_{22}\sin\Psi_B^+ \sin\Phi_A\cos\phi_A\big) =p_{10}'.
    \end{align}
\end{subequations}

Finally, for $\theta_A^{(opt)}=\theta_B^{(opt)}=(2n+1)\pi/2$ with $\phi_A^{(opt)}=0$ and $\phi_B^{(opt)}=(2n+1)\pi/2$ or $\phi_A^{(opt)}=(2n+1)\pi/2$ and $\phi_B^{(opt)}=0$, we have

\begin{subequations}\label{eq:p'_ij-thetaPi/2phi0Pi/2}
    \begin{align}
        p_{00}' =\; & \frac{1}{4} \big( 1 - T_{30}\cos\Phi_A  - T_{03}\cos\Phi_B \notag
        \\
        &\qq{ } +T_{33}\cos\Phi_A\cos\phi_B\big),
        \\
        p_{01}' =\; & \frac{1}{4} \big( 1 - T_{30}\cos\Phi_A  + T_{03}\cos\Phi_B \notag
        \\
        &\qq{ } -T_{33}\cos\Phi_A\cos\phi_B\big),
        \\
        p_{10}' =\; & \frac{1}{4} \big( 1 + T_{30}\cos\Phi_A  - T_{03}\cos\Phi_B \notag
        \\
        &\qq{ } -T_{33}\cos\Phi_A\cos\phi_B\big),
        \\
        p_{11}' =\; & \frac{1}{4} \big( 1 + T_{30}\cos\Phi_A  + T_{03}\cos\Phi_B \notag
        \\
        &\qq{ } +T_{33}\cos\Phi_A\cos\phi_B\big),
    \end{align}
\end{subequations}

 Regarding the marginal probabilities of each subsystem, ${p'_i}^I$, we have
\begin{subequations}
    \begin{align}
        {p'_{\smqty{0\\1}}}^{A} =\;& \frac{1}{2} \qty(1 \mp{}T_{30}\sin\theta_A\cos\Phi_A),
        \\
        {p'_{\smqty{0\\1}}}^{B} =\;& \frac{1}{2} \qty(1 \mp{}T_{03}\sin\theta_B\cos\Phi_B).
    \end{align}
\end{subequations}

Given the above expressions, their maximization lead to the $g_i(T_{i0},T{0i},T_{ii})$ function introduced in Eq. \eqref{eq:Def-gi}.

\bibliographystyle{apsrev}
\bibliography{biblio-qit}

\begin{thebibliography}{85}
\expandafter\ifx\csname natexlab\endcsname\relax\def\natexlab#1{#1}\fi
\expandafter\ifx\csname bibnamefont\endcsname\relax
  \def\bibnamefont#1{#1}\fi
\expandafter\ifx\csname bibfnamefont\endcsname\relax
  \def\bibfnamefont#1{#1}\fi
\expandafter\ifx\csname citenamefont\endcsname\relax
  \def\citenamefont#1{#1}\fi
\expandafter\ifx\csname url\endcsname\relax
  \def\url#1{\texttt{#1}}\fi
\expandafter\ifx\csname urlprefix\endcsname\relax\def\urlprefix{URL }\fi
\providecommand{\bibinfo}[2]{#2}
\providecommand{\eprint}[2][]{\url{#2}}

\bibitem[{\citenamefont{Nielsen and Chuang}(2010)}]{Nielsen-QIT}
\bibinfo{author}{\bibfnamefont{M.}~\bibnamefont{Nielsen}} \bibnamefont{and} \bibinfo{author}{\bibfnamefont{I.}~\bibnamefont{Chuang}}, \emph{\bibinfo{title}{Quantum Computation and Quantum Information: 10th Anniversary Edition}} (\bibinfo{publisher}{Cambridge University Press}, \bibinfo{year}{2010}), ISBN \bibinfo{isbn}{9781139495486}, \urlprefix\url{https://doi.org/10.1017/CBO9780511976667}.

\bibitem[{\citenamefont{Nakahara and Ohmi}(2008)}]{Nakahara-QC}
\bibinfo{author}{\bibfnamefont{M.}~\bibnamefont{Nakahara}} \bibnamefont{and} \bibinfo{author}{\bibfnamefont{T.}~\bibnamefont{Ohmi}}, \emph{\bibinfo{title}{Quantum Computing: From Linear Algebra to Physical Realizations}} (\bibinfo{publisher}{CRC Press}, \bibinfo{year}{2008}), ISBN \bibinfo{isbn}{9781420012293}, \urlprefix\url{https://doi.org/10.1201/9781420012293}.

\bibitem[{\citenamefont{Vedral}(2006)}]{Vedral-QIT}
\bibinfo{author}{\bibfnamefont{V.}~\bibnamefont{Vedral}}, \emph{\bibinfo{title}{Introduction to Quantum Information Science}} (\bibinfo{publisher}{Oxford University Press}, \bibinfo{year}{2006}), ISBN \bibinfo{isbn}{9780199215706}, \urlprefix\url{https://doi.org/10.1093/acprof:oso/9780199215706.001.0001}.

\bibitem[{\citenamefont{Aolita et~al.}(2015)\citenamefont{Aolita, de~Melo, and Davidovich}}]{Aolita2015OpenSysDyn-Review}
\bibinfo{author}{\bibfnamefont{L.}~\bibnamefont{Aolita}}, \bibinfo{author}{\bibfnamefont{F.}~\bibnamefont{de~Melo}}, \bibnamefont{and} \bibinfo{author}{\bibfnamefont{L.}~\bibnamefont{Davidovich}}, \bibinfo{journal}{Reports on Progress in Physics} \textbf{\bibinfo{volume}{78}}, \bibinfo{pages}{042001} (\bibinfo{year}{2015}), ISSN \bibinfo{issn}{0034-4885}, \urlprefix\url{http://dx.doi.org/10.1088/0034-4885/78/4/042001}.

\bibitem[{\citenamefont{Schlosshauer}(2019)}]{Schlosshauer2019QuantumDeco}
\bibinfo{author}{\bibfnamefont{M.}~\bibnamefont{Schlosshauer}}, \bibinfo{journal}{Phys. Rep.} \textbf{\bibinfo{volume}{831}}, \bibinfo{pages}{1} (\bibinfo{year}{2019}), ISSN \bibinfo{issn}{0370-1573}, \urlprefix\url{http://dx.doi.org/10.1016/j.physrep.2019.10.001}.

\bibitem[{\citenamefont{Schlosshauer}(2007)}]{Schlosshauer}
\bibinfo{author}{\bibfnamefont{M.}~\bibnamefont{Schlosshauer}}, \emph{\bibinfo{title}{Decoherence: And the Quantum-To-Classical Transition}}, The Frontiers Collection (\bibinfo{publisher}{Springer-Verlag}, \bibinfo{year}{2007}), ISBN \bibinfo{isbn}{9783540357735}, \urlprefix\url{https://doi.org/10.1007/978-3-540-35775-9}.

\bibitem[{\citenamefont{Campaioli et~al.}(2024)\citenamefont{Campaioli, Cole, and Hapuarachchi}}]{Campaioli2024QuantumMasterEq}
\bibinfo{author}{\bibfnamefont{F.}~\bibnamefont{Campaioli}}, \bibinfo{author}{\bibfnamefont{J.~H.} \bibnamefont{Cole}}, \bibnamefont{and} \bibinfo{author}{\bibfnamefont{H.}~\bibnamefont{Hapuarachchi}}, \bibinfo{journal}{PRX Quantum} \textbf{\bibinfo{volume}{5}} (\bibinfo{year}{2024}), ISSN \bibinfo{issn}{2691-3399}, \urlprefix\url{http://dx.doi.org/10.1103/PRXQuantum.5.020202}.

\bibitem[{\citenamefont{Gorini et~al.}(1976)\citenamefont{Gorini, Kossakowski, and Sudarshan}}]{Gorini1976-GKS_Eq}
\bibinfo{author}{\bibfnamefont{V.}~\bibnamefont{Gorini}}, \bibinfo{author}{\bibfnamefont{A.}~\bibnamefont{Kossakowski}}, \bibnamefont{and} \bibinfo{author}{\bibfnamefont{E.~C.~G.} \bibnamefont{Sudarshan}}, \bibinfo{journal}{J. Math. Phys.} \textbf{\bibinfo{volume}{17}}, \bibinfo{pages}{821} (\bibinfo{year}{1976}), ISSN \bibinfo{issn}{0022-2488}, \urlprefix\url{http://dx.doi.org/10.1063/1.522979}.

\bibitem[{\citenamefont{Lindblad}(1976)}]{Lindblad1976-MasterEq}
\bibinfo{author}{\bibfnamefont{G.}~\bibnamefont{Lindblad}}, \bibinfo{journal}{Commun. Math. Phys.} \textbf{\bibinfo{volume}{48}}, \bibinfo{pages}{119} (\bibinfo{year}{1976}), ISSN \bibinfo{issn}{0010-3616}, \urlprefix\url{http://dx.doi.org/10.1007/BF01608499}.

\bibitem[{\citenamefont{Manzano}(2020)}]{Manzano2020-GKSL_Eq}
\bibinfo{author}{\bibfnamefont{D.}~\bibnamefont{Manzano}}, \bibinfo{journal}{AIP Adv.} \textbf{\bibinfo{volume}{10}} (\bibinfo{year}{2020}), ISSN \bibinfo{issn}{2158-3226}, \urlprefix\url{http://dx.doi.org/10.1063/1.5115323}.

\bibitem[{\citenamefont{Havel}(2003)}]{Havel2003-KrausLind}
\bibinfo{author}{\bibfnamefont{T.~F.} \bibnamefont{Havel}}, \bibinfo{journal}{J. Math. Phys.} \textbf{\bibinfo{volume}{44}}, \bibinfo{pages}{534} (\bibinfo{year}{2003}), ISSN \bibinfo{issn}{0022-2488}, \urlprefix\url{http://dx.doi.org/10.1063/1.1518555}.

\bibitem[{\citenamefont{Nakazato et~al.}(2006)\citenamefont{Nakazato, Hida, Yuasa, Militello, Napoli, and Messina}}]{Nakazato2006-KrausLind}
\bibinfo{author}{\bibfnamefont{H.}~\bibnamefont{Nakazato}}, \bibinfo{author}{\bibfnamefont{Y.}~\bibnamefont{Hida}}, \bibinfo{author}{\bibfnamefont{K.}~\bibnamefont{Yuasa}}, \bibinfo{author}{\bibfnamefont{B.}~\bibnamefont{Militello}}, \bibinfo{author}{\bibfnamefont{A.}~\bibnamefont{Napoli}}, \bibnamefont{and} \bibinfo{author}{\bibfnamefont{A.}~\bibnamefont{Messina}}, \bibinfo{journal}{Phys. Rev. A} \textbf{\bibinfo{volume}{74}} (\bibinfo{year}{2006}), ISSN \bibinfo{issn}{1050-2947}, \urlprefix\url{http://dx.doi.org/10.1103/PhysRevA.74.062113}.

\bibitem[{\citenamefont{Andersson et~al.}(2007)\citenamefont{Andersson, Cresser, and Hall}}]{Andersson2007-KrausLind}
\bibinfo{author}{\bibfnamefont{E.}~\bibnamefont{Andersson}}, \bibinfo{author}{\bibfnamefont{J.~D.} \bibnamefont{Cresser}}, \bibnamefont{and} \bibinfo{author}{\bibfnamefont{M.~J.~W.} \bibnamefont{Hall}}, \bibinfo{journal}{J. Mod. Optics} \textbf{\bibinfo{volume}{54}}, \bibinfo{pages}{1695} (\bibinfo{year}{2007}), ISSN \bibinfo{issn}{0950-0340}, \urlprefix\url{http://dx.doi.org/10.1080/09500340701352581}.

\bibitem[{\citenamefont{Kraus}(1971)}]{Kraus-Article}
\bibinfo{author}{\bibfnamefont{K.}~\bibnamefont{Kraus}}, \bibinfo{journal}{Ann. Physics} \textbf{\bibinfo{volume}{64}}, \bibinfo{pages}{311} (\bibinfo{year}{1971}), ISSN \bibinfo{issn}{0003-4916}, \urlprefix\url{http://dx.doi.org/10.1016/0003-4916(71)90108-4}.

\bibitem[{\citenamefont{Kraus}(1983)}]{Kraus-Lectures}
\bibinfo{author}{\bibfnamefont{K.}~\bibnamefont{Kraus}}, \emph{\bibinfo{title}{States, effects, and operations: Fundamental notions of quantum theory. Lectures in Mathematical Physics at the University of Texas at Austin}}, Lecture notes in physics (\bibinfo{publisher}{Springer-Verlag}, \bibinfo{year}{1983}), ISBN \bibinfo{isbn}{9780387127323}, \urlprefix\url{https://doi.org/10.1007/3-540-12732-1}.

\bibitem[{\citenamefont{Gardiner and Zoller}(2004)}]{GardinerZoller-QuantumNoise}
\bibinfo{author}{\bibfnamefont{C.}~\bibnamefont{Gardiner}} \bibnamefont{and} \bibinfo{author}{\bibfnamefont{P.}~\bibnamefont{Zoller}}, \emph{\bibinfo{title}{Quantum Noise: A Handbook of Markovian and Non-Markovian Quantum Stochastic Methods with Applications to Quantum Optics}}, Springer Series in Synergetics (\bibinfo{publisher}{Springer-Verlag}, \bibinfo{year}{2004}), ISBN \bibinfo{isbn}{9783540223016}, \urlprefix\url{https://link.springer.com/book/9783540223016}.

\bibitem[{\citenamefont{Daffer et~al.}(2004)\citenamefont{Daffer, W{\' o}dkiewicz, Cresser, and McIver}}]{Daffer2004-QuantumRTN}
\bibinfo{author}{\bibfnamefont{S.}~\bibnamefont{Daffer}}, \bibinfo{author}{\bibfnamefont{K.}~\bibnamefont{W{\' o}dkiewicz}}, \bibinfo{author}{\bibfnamefont{J.~D.} \bibnamefont{Cresser}}, \bibnamefont{and} \bibinfo{author}{\bibfnamefont{J.~K.} \bibnamefont{McIver}}, \bibinfo{journal}{Phys. Rev. A} \textbf{\bibinfo{volume}{70}} (\bibinfo{year}{2004}), ISSN \bibinfo{issn}{1050-2947}, \urlprefix\url{http://dx.doi.org/10.1103/PhysRevA.70.010304}.

\bibitem[{\citenamefont{Kumar et~al.}(2018)\citenamefont{Kumar, Banerjee, Srikanth, Jagadish, and Petruccione}}]{Kumar2018-NonMarkovianGameTheory}
\bibinfo{author}{\bibfnamefont{N.~P.} \bibnamefont{Kumar}}, \bibinfo{author}{\bibfnamefont{S.}~\bibnamefont{Banerjee}}, \bibinfo{author}{\bibfnamefont{R.}~\bibnamefont{Srikanth}}, \bibinfo{author}{\bibfnamefont{V.}~\bibnamefont{Jagadish}}, \bibnamefont{and} \bibinfo{author}{\bibfnamefont{F.}~\bibnamefont{Petruccione}}, \bibinfo{journal}{Open Syst. Inf. Dyn.} \textbf{\bibinfo{volume}{25}}, \bibinfo{pages}{1850014} (\bibinfo{year}{2018}), ISSN \bibinfo{issn}{1230-1612}, \urlprefix\url{http://dx.doi.org/10.1142/S1230161218500142}.

\bibitem[{\citenamefont{Thapliyal et~al.}(2017)\citenamefont{Thapliyal, Pathak, and Banerjee}}]{Thapliyal2017-qCrypto-NonMarkovChannel}
\bibinfo{author}{\bibfnamefont{K.}~\bibnamefont{Thapliyal}}, \bibinfo{author}{\bibfnamefont{A.}~\bibnamefont{Pathak}}, \bibnamefont{and} \bibinfo{author}{\bibfnamefont{S.}~\bibnamefont{Banerjee}}, \bibinfo{journal}{Quantum Inf. Process.} \textbf{\bibinfo{volume}{16}} (\bibinfo{year}{2017}), ISSN \bibinfo{issn}{1570-0755}, \urlprefix\url{http://dx.doi.org/10.1007/s11128-017-1567-1}.

\bibitem[{\citenamefont{Naikoo et~al.}(2019)\citenamefont{Naikoo, Dutta, and Banerjee}}]{Naikoo2019-QI_NonMarkovEvol}
\bibinfo{author}{\bibfnamefont{J.}~\bibnamefont{Naikoo}}, \bibinfo{author}{\bibfnamefont{S.}~\bibnamefont{Dutta}}, \bibnamefont{and} \bibinfo{author}{\bibfnamefont{S.}~\bibnamefont{Banerjee}}, \bibinfo{journal}{Phys. Rev. A} \textbf{\bibinfo{volume}{99}} (\bibinfo{year}{2019}), ISSN \bibinfo{issn}{2469-9926}, \urlprefix\url{http://dx.doi.org/10.1103/PhysRevA.99.042128}.

\bibitem[{\citenamefont{Mishra et~al.}(2022)\citenamefont{Mishra, Thapliyal, and Pathak}}]{Mishra2022-QIP-NonMark-Xstates}
\bibinfo{author}{\bibfnamefont{S.}~\bibnamefont{Mishra}}, \bibinfo{author}{\bibfnamefont{K.}~\bibnamefont{Thapliyal}}, \bibnamefont{and} \bibinfo{author}{\bibfnamefont{A.}~\bibnamefont{Pathak}}, \bibinfo{journal}{Quantum Inf. Process.} \textbf{\bibinfo{volume}{21}} (\bibinfo{year}{2022}), ISSN \bibinfo{issn}{1570-0755}, \urlprefix\url{http://dx.doi.org/10.1007/s11128-021-03408-2}.

\bibitem[{\citenamefont{Sabale et~al.}(2024)\citenamefont{Sabale, Dash, Kumar, and Banerjee}}]{Sabale2024Dchannels-RTN-OUN}
\bibinfo{author}{\bibfnamefont{V.~B.} \bibnamefont{Sabale}}, \bibinfo{author}{\bibfnamefont{N.~R.} \bibnamefont{Dash}}, \bibinfo{author}{\bibfnamefont{A.}~\bibnamefont{Kumar}}, \bibnamefont{and} \bibinfo{author}{\bibfnamefont{S.}~\bibnamefont{Banerjee}}, \bibinfo{journal}{Annalen der Physik} \textbf{\bibinfo{volume}{536}} (\bibinfo{year}{2024}), ISSN \bibinfo{issn}{0003-3804}, \urlprefix\url{http://dx.doi.org/10.1002/andp.202400151}.

\bibitem[{\citenamefont{Aiache et~al.}(2025)\citenamefont{Aiache, Seida, El~Anouz, and El~Allati}}]{Aiache2025-Dephasing}
\bibinfo{author}{\bibfnamefont{Y.}~\bibnamefont{Aiache}}, \bibinfo{author}{\bibfnamefont{C.}~\bibnamefont{Seida}}, \bibinfo{author}{\bibfnamefont{K.}~\bibnamefont{El~Anouz}}, \bibnamefont{and} \bibinfo{author}{\bibfnamefont{A.}~\bibnamefont{El~Allati}}, \bibinfo{journal}{AVS Quantum Sci.} \textbf{\bibinfo{volume}{7}} (\bibinfo{year}{2025}), ISSN \bibinfo{issn}{2639-0213}, \urlprefix\url{http://dx.doi.org/10.1116/5.0268432}.

\bibitem[{\citenamefont{Benedetti et~al.}(2013)\citenamefont{Benedetti, Buscemi, Bordone, and Paris}}]{Benedetti2013-ColoredNoise}
\bibinfo{author}{\bibfnamefont{C.}~\bibnamefont{Benedetti}}, \bibinfo{author}{\bibfnamefont{F.}~\bibnamefont{Buscemi}}, \bibinfo{author}{\bibfnamefont{P.}~\bibnamefont{Bordone}}, \bibnamefont{and} \bibinfo{author}{\bibfnamefont{M.~G.~A.} \bibnamefont{Paris}}, \bibinfo{journal}{Phys. Rev. A} \textbf{\bibinfo{volume}{87}} (\bibinfo{year}{2013}), ISSN \bibinfo{issn}{1050-2947}, \urlprefix\url{http://dx.doi.org/10.1103/PhysRevA.87.052328}.

\bibitem[{\citenamefont{Benedetti et~al.}(2014)\citenamefont{Benedetti, Buscemi, Bordone, and Paris}}]{Benedetti2014-RTN&ColoredNoise}
\bibinfo{author}{\bibfnamefont{C.}~\bibnamefont{Benedetti}}, \bibinfo{author}{\bibfnamefont{F.}~\bibnamefont{Buscemi}}, \bibinfo{author}{\bibfnamefont{P.}~\bibnamefont{Bordone}}, \bibnamefont{and} \bibinfo{author}{\bibfnamefont{M.~G.~A.} \bibnamefont{Paris}}, \bibinfo{journal}{Phys. Rev. A} \textbf{\bibinfo{volume}{89}} (\bibinfo{year}{2014}), ISSN \bibinfo{issn}{1050-2947}, \urlprefix\url{http://dx.doi.org/10.1103/PhysRevA.89.032114}.

\bibitem[{\citenamefont{Uhlenbeck and Ornstein}(1930)}]{Uhlenbeck1930O-OU-original}
\bibinfo{author}{\bibfnamefont{G.~E.} \bibnamefont{Uhlenbeck}} \bibnamefont{and} \bibinfo{author}{\bibfnamefont{L.~S.} \bibnamefont{Ornstein}}, \bibinfo{journal}{Phys. Rev.} \textbf{\bibinfo{volume}{36}}, \bibinfo{pages}{823} (\bibinfo{year}{1930}), ISSN \bibinfo{issn}{0031-899X}, \urlprefix\url{http://dx.doi.org/10.1103/PhysRev.36.823}.

\bibitem[{\citenamefont{Maller et~al.}(2009)\citenamefont{Maller, M{\" u}ller, and Szimayer}}]{Maller2009-Gen_OrnsteinUhlenbeck}
\bibinfo{author}{\bibfnamefont{R.~A.} \bibnamefont{Maller}}, \bibinfo{author}{\bibfnamefont{G.}~\bibnamefont{M{\" u}ller}}, \bibnamefont{and} \bibinfo{author}{\bibfnamefont{A.}~\bibnamefont{Szimayer}}, \emph{\bibinfo{title}{Ornstein--{Uhlenbeck} {Processes} and {Extensions}}} (\bibinfo{publisher}{Springer-Verlag}, \bibinfo{year}{2009}), pp. \bibinfo{pages}{421--437}, ISBN \bibinfo{isbn}{9783540712961}, \urlprefix\url{http://dx.doi.org/10.1007/978-3-540-71297-8_18}.

\bibitem[{\citenamefont{Tiersch et~al.}(2013)\citenamefont{Tiersch, de~Melo, and Buchleitner}}]{Tiersch2013Universality}
\bibinfo{author}{\bibfnamefont{M.}~\bibnamefont{Tiersch}}, \bibinfo{author}{\bibfnamefont{F.}~\bibnamefont{de~Melo}}, \bibnamefont{and} \bibinfo{author}{\bibfnamefont{A.}~\bibnamefont{Buchleitner}}, \bibinfo{journal}{J. Phy. A Math. Theor.} \textbf{\bibinfo{volume}{46}}, \bibinfo{pages}{085301} (\bibinfo{year}{2013}), ISSN \bibinfo{issn}{1751-8113}, \urlprefix\url{http://dx.doi.org/10.1088/1751-8113/46/8/085301}.

\bibitem[{\citenamefont{Yu and Eberly}(2007)}]{EstadosX}
\bibinfo{author}{\bibfnamefont{T.}~\bibnamefont{Yu}} \bibnamefont{and} \bibinfo{author}{\bibfnamefont{J.~H.} \bibnamefont{Eberly}}, \bibinfo{journal}{Quantum Inf. Comput.} \textbf{\bibinfo{volume}{7}}, \bibinfo{pages}{459} (\bibinfo{year}{2007}), \urlprefix\url{https://dl.acm.org/doi/10.5555/2011832.2011835}.

\bibitem[{\citenamefont{Quesada et~al.}(2012)\citenamefont{Quesada, Al-Qasimi, and James}}]{Quesada-XStates}
\bibinfo{author}{\bibfnamefont{N.}~\bibnamefont{Quesada}}, \bibinfo{author}{\bibfnamefont{A.}~\bibnamefont{Al-Qasimi}}, \bibnamefont{and} \bibinfo{author}{\bibfnamefont{D.}~\bibnamefont{James}}, \bibinfo{journal}{J. Mod. Opt.} \textbf{\bibinfo{volume}{59}}, \bibinfo{pages}{1322} (\bibinfo{year}{2012}), \bibinfo{note}{arXiv:1207.3689}, \urlprefix\url{http://dx.doi.org/10.1080/09500340.2012.713130}.

\bibitem[{\citenamefont{Rau}(2009)}]{Rau_2009-Xstates_algebra}
\bibinfo{author}{\bibfnamefont{A.~R.~P.} \bibnamefont{Rau}}, \bibinfo{journal}{J. Phys. A: Math. Gen.} \textbf{\bibinfo{volume}{42}}, \bibinfo{pages}{412002} (\bibinfo{year}{2009}), \urlprefix\url{https://doi.org/10.1088/1751-8113/42/41/412002}.

\bibitem[{\citenamefont{Mendon{\c{c}}a et~al.}(2014)\citenamefont{Mendon{\c{c}}a, Marchiolli, and Galetti}}]{XStates-Entanglement}
\bibinfo{author}{\bibfnamefont{P.~E. M.~F.} \bibnamefont{Mendon{\c{c}}a}}, \bibinfo{author}{\bibfnamefont{M.~A.} \bibnamefont{Marchiolli}}, \bibnamefont{and} \bibinfo{author}{\bibfnamefont{D.}~\bibnamefont{Galetti}}, \bibinfo{journal}{Ann. Physics} \textbf{\bibinfo{volume}{351}}, \bibinfo{pages}{79} (\bibinfo{year}{2014}), \bibinfo{note}{arXiv:1407.3021}, \urlprefix\url{http://dx.doi.org/10.1016/j.aop.2014.08.017}.

\bibitem[{\citenamefont{Hedemann}(2018)}]{Hedemann-XStates}
\bibinfo{author}{\bibfnamefont{S.}~\bibnamefont{Hedemann}}, \bibinfo{journal}{Quantum Inf. Process.} \textbf{\bibinfo{volume}{17}}, \bibinfo{pages}{293} (\bibinfo{year}{2018}), \bibinfo{note}{arXiv:1802.03038}.

\bibitem[{\citenamefont{Zhou et~al.}(2012)\citenamefont{Zhou, Zhang, Fei, Jing, and Li-Jost}}]{Zhou-CanonicalXstates}
\bibinfo{author}{\bibfnamefont{C.}~\bibnamefont{Zhou}}, \bibinfo{author}{\bibfnamefont{T.-G.} \bibnamefont{Zhang}}, \bibinfo{author}{\bibfnamefont{S.-M.} \bibnamefont{Fei}}, \bibinfo{author}{\bibfnamefont{N.}~\bibnamefont{Jing}}, \bibnamefont{and} \bibinfo{author}{\bibfnamefont{X.}~\bibnamefont{Li-Jost}}, \bibinfo{journal}{Phys. Rev. A} \textbf{\bibinfo{volume}{86}}, \bibinfo{pages}{010303} (\bibinfo{year}{2012}), \urlprefix\url{https://link.aps.org/doi/10.1103/PhysRevA.86.010303}.

\bibitem[{\citenamefont{Wootters}(1998)}]{Wooters_Concurrence}
\bibinfo{author}{\bibfnamefont{W.}~\bibnamefont{Wootters}}, \bibinfo{journal}{Phys. Rev. Lett.} \textbf{\bibinfo{volume}{80}}, \bibinfo{pages}{2245} (\bibinfo{year}{1998}), \bibinfo{note}{arXiv:quantph/9709029}.

\bibitem[{\citenamefont{Horodecki et~al.}(2009)\citenamefont{Horodecki, Horodecki, Horodecki, and Horodecki}}]{Horodecki-Ent}
\bibinfo{author}{\bibfnamefont{R.}~\bibnamefont{Horodecki}}, \bibinfo{author}{\bibfnamefont{P.}~\bibnamefont{Horodecki}}, \bibinfo{author}{\bibfnamefont{M.}~\bibnamefont{Horodecki}}, \bibnamefont{and} \bibinfo{author}{\bibfnamefont{K.}~\bibnamefont{Horodecki}}, \bibinfo{journal}{Rev. Mod. Phys.} \textbf{\bibinfo{volume}{81}}, \bibinfo{pages}{865} (\bibinfo{year}{2009}), \bibinfo{note}{arXiv:quant-ph/0702225}, \urlprefix\url{http://dx.doi.org/10.1103/RevModPhys.81.865}.

\bibitem[{\citenamefont{Tiersch et~al.}(2009)\citenamefont{Tiersch, de~Melo, Konrad, and Buchleitner}}]{Tiersch2009EquationEntanglement}
\bibinfo{author}{\bibfnamefont{M.}~\bibnamefont{Tiersch}}, \bibinfo{author}{\bibfnamefont{F.}~\bibnamefont{de~Melo}}, \bibinfo{author}{\bibfnamefont{T.}~\bibnamefont{Konrad}}, \bibnamefont{and} \bibinfo{author}{\bibfnamefont{A.}~\bibnamefont{Buchleitner}}, \bibinfo{journal}{Quantum Information Processing} \textbf{\bibinfo{volume}{8}}, \bibinfo{pages}{523} (\bibinfo{year}{2009}), ISSN \bibinfo{issn}{1570-0755}, \urlprefix\url{http://dx.doi.org/10.1007/s11128-009-0139-4}.

\bibitem[{\citenamefont{Ollivier and Zurek}(2001)}]{qDiscord-Olliver}
\bibinfo{author}{\bibfnamefont{H.}~\bibnamefont{Ollivier}} \bibnamefont{and} \bibinfo{author}{\bibfnamefont{W.~H.} \bibnamefont{Zurek}}, \bibinfo{journal}{Phys. Rev. Lett.} \textbf{\bibinfo{volume}{88}} (\bibinfo{year}{2001}), ISSN \bibinfo{issn}{0031-9007}, \urlprefix\url{http://dx.doi.org/10.1103/PhysRevLett.88.017901}.

\bibitem[{\citenamefont{Henderson and Vedral}(2001)}]{qDiscord-Henderson}
\bibinfo{author}{\bibfnamefont{L.}~\bibnamefont{Henderson}} \bibnamefont{and} \bibinfo{author}{\bibfnamefont{V.}~\bibnamefont{Vedral}}, \bibinfo{journal}{J. Phys. A: Math. Gen.} \textbf{\bibinfo{volume}{34}}, \bibinfo{pages}{6899} (\bibinfo{year}{2001}), ISSN \bibinfo{issn}{0305-4470}, \urlprefix\url{http://dx.doi.org/10.1088/0305-4470/34/35/315}.

\bibitem[{\citenamefont{Modi et~al.}(2012)\citenamefont{Modi, Brodutch, Cable, Paterek, and Vedral}}]{Modi-qDiscord}
\bibinfo{author}{\bibfnamefont{K.}~\bibnamefont{Modi}}, \bibinfo{author}{\bibfnamefont{A.}~\bibnamefont{Brodutch}}, \bibinfo{author}{\bibfnamefont{H.}~\bibnamefont{Cable}}, \bibinfo{author}{\bibfnamefont{T.}~\bibnamefont{Paterek}}, \bibnamefont{and} \bibinfo{author}{\bibfnamefont{V.}~\bibnamefont{Vedral}}, \bibinfo{journal}{Rev. Mod. Phys.} \textbf{\bibinfo{volume}{84}}, \bibinfo{pages}{1655} (\bibinfo{year}{2012}), \bibinfo{note}{arXiv:1112.6238}.

\bibitem[{\citenamefont{Braunstein and Caves}(1994)}]{Braunstein1994-qFisherInfo}
\bibinfo{author}{\bibfnamefont{S.~L.} \bibnamefont{Braunstein}} \bibnamefont{and} \bibinfo{author}{\bibfnamefont{C.~M.} \bibnamefont{Caves}}, \bibinfo{journal}{Phys. Rev. Lett.} \textbf{\bibinfo{volume}{72}}, \bibinfo{pages}{3439} (\bibinfo{year}{1994}), ISSN \bibinfo{issn}{0031-9007}, \urlprefix\url{http://dx.doi.org/10.1103/PhysRevLett.72.3439}.

\bibitem[{\citenamefont{Braunstein et~al.}(1996)\citenamefont{Braunstein, Caves, and Milburn}}]{Braunstein1996-qFisherInfo}
\bibinfo{author}{\bibfnamefont{S.~L.} \bibnamefont{Braunstein}}, \bibinfo{author}{\bibfnamefont{C.~M.} \bibnamefont{Caves}}, \bibnamefont{and} \bibinfo{author}{\bibfnamefont{G.}~\bibnamefont{Milburn}}, \bibinfo{journal}{Ann. Phys. (N. Y.)} \textbf{\bibinfo{volume}{247}}, \bibinfo{pages}{135} (\bibinfo{year}{1996}), ISSN \bibinfo{issn}{0003-4916}, \urlprefix\url{http://dx.doi.org/10.1006/aphy.1996.0040}.

\bibitem[{\citenamefont{Shahbeigi and Akhtarshenas}(2018)}]{Shahbeigi2018-QuantumnessOfChannels}
\bibinfo{author}{\bibfnamefont{F.}~\bibnamefont{Shahbeigi}} \bibnamefont{and} \bibinfo{author}{\bibfnamefont{S.~J.} \bibnamefont{Akhtarshenas}}, \bibinfo{journal}{Phys. Rev. A} \textbf{\bibinfo{volume}{98}} (\bibinfo{year}{2018}), ISSN \bibinfo{issn}{2469-9926}, \urlprefix\url{http://dx.doi.org/10.1103/PhysRevA.98.042313}.

\bibitem[{\citenamefont{Holevo}(1973)}]{Holevo}
\bibinfo{author}{\bibfnamefont{A.~S.} \bibnamefont{Holevo}}, \bibinfo{journal}{Probl. Inf. Transm.} \textbf{\bibinfo{volume}{9}}, \bibinfo{pages}{177} (\bibinfo{year}{1973}), \urlprefix\url{https://www.mathnet.ru/eng/ppi903}.

\bibitem[{\citenamefont{Wu et~al.}(2009)\citenamefont{Wu, Poulsen, and M{\o}lmer}}]{Wu_AMID}
\bibinfo{author}{\bibfnamefont{S.}~\bibnamefont{Wu}}, \bibinfo{author}{\bibfnamefont{U.}~\bibnamefont{Poulsen}}, \bibnamefont{and} \bibinfo{author}{\bibfnamefont{K.}~\bibnamefont{M{\o}lmer}}, \bibinfo{journal}{Phys. Rev. A} \textbf{\bibinfo{volume}{80}}, \bibinfo{pages}{032319} (\bibinfo{year}{2009}), ISSN \bibinfo{issn}{1094-1622}, \urlprefix\url{https://doi.org/10.1103/PhysRevA.80.032319}.

\bibitem[{\citenamefont{Wu et~al.}(2015)\citenamefont{Wu, Ma, Chen, and Xia}}]{Wu_ComplementaryBases}
\bibinfo{author}{\bibfnamefont{S.}~\bibnamefont{Wu}}, \bibinfo{author}{\bibfnamefont{Z.}~\bibnamefont{Ma}}, \bibinfo{author}{\bibfnamefont{Z.}~\bibnamefont{Chen}}, \bibnamefont{and} \bibinfo{author}{\bibfnamefont{S.}~\bibnamefont{Xia}}, \bibinfo{journal}{Sci. Rep.} \textbf{\bibinfo{volume}{4}}, \bibinfo{pages}{4036} (\bibinfo{year}{2015}), \bibinfo{note}{arXiv:1301.6838}, \urlprefix\url{http://dx.doi.org/10.1038/srep04036}.

\bibitem[{\citenamefont{Schwinger}(1960)}]{Schwinger_MUB}
\bibinfo{author}{\bibfnamefont{J.}~\bibnamefont{Schwinger}}, \bibinfo{journal}{Proc. Natl. Acad. Sci. U.S.A.} \textbf{\bibinfo{volume}{46}}, \bibinfo{pages}{570} (\bibinfo{year}{1960}), \eprint{https://www.pnas.org/doi/pdf/10.1073/pnas.46.4.570}, \urlprefix\url{https://www.pnas.org/doi/abs/10.1073/pnas.46.4.570}.

\bibitem[{\citenamefont{Bengtsson}(2007)}]{Bengtsson2007-MUB}
\bibinfo{author}{\bibfnamefont{I.}~\bibnamefont{Bengtsson}}, \bibinfo{journal}{AIP Conf. Proc.} \textbf{\bibinfo{volume}{889}}, \bibinfo{pages}{40} (\bibinfo{year}{2007}), ISSN \bibinfo{issn}{0094-243X}, \urlprefix\url{http://dx.doi.org/10.1063/1.2713445}.

\bibitem[{\citenamefont{Durt et~al.}(2010)\citenamefont{Durt, Englert, Bengtsson, and Życzkowski}}]{DURT2010-MUB}
\bibinfo{author}{\bibfnamefont{T.}~\bibnamefont{Durt}}, \bibinfo{author}{\bibfnamefont{B.-G.} \bibnamefont{Englert}}, \bibinfo{author}{\bibfnamefont{I.}~\bibnamefont{Bengtsson}}, \bibnamefont{and} \bibinfo{author}{\bibfnamefont{K.}~\bibnamefont{Życzkowski}}, \bibinfo{journal}{Int. J. Quantum Inf.} \textbf{\bibinfo{volume}{8}}, \bibinfo{pages}{535} (\bibinfo{year}{2010}), ISSN \bibinfo{issn}{0219-7499}, \urlprefix\url{http://dx.doi.org/10.1142/S0219749910006502}.

\bibitem[{\citenamefont{Mundarain and Ladr{\' o}n~de Guevara}(2015)}]{LAQC}
\bibinfo{author}{\bibfnamefont{D.~F.} \bibnamefont{Mundarain}} \bibnamefont{and} \bibinfo{author}{\bibfnamefont{M.~L.} \bibnamefont{Ladr{\' o}n~de Guevara}}, \bibinfo{journal}{Quantum Inf. Process.} \textbf{\bibinfo{volume}{14}}, \bibinfo{pages}{4493} (\bibinfo{year}{2015}), ISSN \bibinfo{issn}{1570-0755}, \urlprefix\url{http://dx.doi.org/10.1007/s11128-015-1139-1}.

\bibitem[{\citenamefont{Piani et~al.}(2011)\citenamefont{Piani, Gharibian, Adesso, Calsamiglia, Horodecki, and Winter}}]{Piani_2011-EntActiv}
\bibinfo{author}{\bibfnamefont{M.}~\bibnamefont{Piani}}, \bibinfo{author}{\bibfnamefont{S.}~\bibnamefont{Gharibian}}, \bibinfo{author}{\bibfnamefont{G.}~\bibnamefont{Adesso}}, \bibinfo{author}{\bibfnamefont{J.}~\bibnamefont{Calsamiglia}}, \bibinfo{author}{\bibfnamefont{P.}~\bibnamefont{Horodecki}}, \bibnamefont{and} \bibinfo{author}{\bibfnamefont{A.}~\bibnamefont{Winter}}, \bibinfo{journal}{Physical Review Letters} \textbf{\bibinfo{volume}{106}} (\bibinfo{year}{2011}), ISSN \bibinfo{issn}{0031-9007}, \urlprefix\url{http://dx.doi.org/10.1103/PhysRevLett.106.220403}.

\bibitem[{\citenamefont{Mazzola and Paternostro}(2011)}]{EntAct_Exp_Mazzola2011}
\bibinfo{author}{\bibfnamefont{L.}~\bibnamefont{Mazzola}} \bibnamefont{and} \bibinfo{author}{\bibfnamefont{M.}~\bibnamefont{Paternostro}}, \bibinfo{journal}{Scientific Reports} \textbf{\bibinfo{volume}{1}}, \bibinfo{pages}{199} (\bibinfo{year}{2011}), ISSN \bibinfo{issn}{2045-2322}, \urlprefix\url{https://doi.org/10.1038/srep00199}.

\bibitem[{\citenamefont{Adesso et~al.}(2014)\citenamefont{Adesso, D'Ambrosio, Nagali, Piani, and Sciarrino}}]{EntAct_Exp_Adesso014}
\bibinfo{author}{\bibfnamefont{G.}~\bibnamefont{Adesso}}, \bibinfo{author}{\bibfnamefont{V.}~\bibnamefont{D'Ambrosio}}, \bibinfo{author}{\bibfnamefont{E.}~\bibnamefont{Nagali}}, \bibinfo{author}{\bibfnamefont{M.}~\bibnamefont{Piani}}, \bibnamefont{and} \bibinfo{author}{\bibfnamefont{F.}~\bibnamefont{Sciarrino}}, \bibinfo{journal}{Phys. Rev. Lett.} \textbf{\bibinfo{volume}{112}}, \bibinfo{pages}{140501} (\bibinfo{year}{2014}), \urlprefix\url{https://link.aps.org/doi/10.1103/PhysRevLett.112.140501}.

\bibitem[{\citenamefont{GHARIBIAN et~al.}(2011)\citenamefont{GHARIBIAN, PIANI, ADESSO, CALSAMIGLIA, and HORODECKI}}]{EntAct_Swapping_Gharibian2011}
\bibinfo{author}{\bibfnamefont{S.}~\bibnamefont{GHARIBIAN}}, \bibinfo{author}{\bibfnamefont{M.}~\bibnamefont{PIANI}}, \bibinfo{author}{\bibfnamefont{G.}~\bibnamefont{ADESSO}}, \bibinfo{author}{\bibfnamefont{J.}~\bibnamefont{CALSAMIGLIA}}, \bibnamefont{and} \bibinfo{author}{\bibfnamefont{P.}~\bibnamefont{HORODECKI}}, \bibinfo{journal}{International Journal of Quantum Information} \textbf{\bibinfo{volume}{09}}, \bibinfo{pages}{1701} (\bibinfo{year}{2011}), \eprint{https://doi.org/10.1142/S0219749911008258}, \urlprefix\url{https://doi.org/10.1142/S0219749911008258}.

\bibitem[{\citenamefont{Albrecht~Q. et~al.}(2018)\citenamefont{Albrecht~Q., Caicedo~S., and Mundarain}}]{LAQC_BD}
\bibinfo{author}{\bibfnamefont{H.}~\bibnamefont{Albrecht~Q.}}, \bibinfo{author}{\bibfnamefont{M.}~\bibnamefont{Caicedo~S.}}, \bibnamefont{and} \bibinfo{author}{\bibfnamefont{D.}~\bibnamefont{Mundarain}}, \bibinfo{journal}{Rev. Mex. F{\'{\i}}s.} \textbf{\bibinfo{volume}{64}}, \bibinfo{pages}{662} (\bibinfo{year}{2018}), \urlprefix\url{http://dx.doi.org/10.31349/RevMexFis.64.662}.

\bibitem[{\citenamefont{Bellorin and Albrecht~Q.}(2021)}]{LAQC_BD-Err}
\bibinfo{author}{\bibfnamefont{D.}~\bibnamefont{Bellorin}} \bibnamefont{and} \bibinfo{author}{\bibfnamefont{H.}~\bibnamefont{Albrecht~Q.}}, \bibinfo{journal}{Rev. Mex. F{\'{\i}}s.} \textbf{\bibinfo{volume}{67}}, \bibinfo{pages}{052301} (\bibinfo{year}{2021}), \bibinfo{note}{arXiv:2105.15166}, \urlprefix\url{https://doi.org/10.31349/RevMexFis.67.052301}.

\bibitem[{\citenamefont{Albrecht~Q. et~al.}(2022)\citenamefont{Albrecht~Q., Bellorin, and Mundarain}}]{LAQC_Xstates-sym}
\bibinfo{author}{\bibfnamefont{H.}~\bibnamefont{Albrecht~Q.}}, \bibinfo{author}{\bibfnamefont{D.}~\bibnamefont{Bellorin}}, \bibnamefont{and} \bibinfo{author}{\bibfnamefont{D.~F.} \bibnamefont{Mundarain}}, \bibinfo{journal}{Int. J. Mod. Phys. B} \textbf{\bibinfo{volume}{36}}, \bibinfo{pages}{22500990} (\bibinfo{year}{2022}), \bibinfo{note}{arXiv:2107.00158}, \urlprefix\url{https://www.worldscientific.com/doi/10.1142/S0217979222500990}.

\bibitem[{\citenamefont{Bellorin et~al.}(2022)\citenamefont{Bellorin, Albrecht~Q., and Mundarain}}]{LAQC_Xstates-no_sym}
\bibinfo{author}{\bibfnamefont{D.}~\bibnamefont{Bellorin}}, \bibinfo{author}{\bibfnamefont{H.}~\bibnamefont{Albrecht~Q.}}, \bibnamefont{and} \bibinfo{author}{\bibfnamefont{D.~F.} \bibnamefont{Mundarain}}, \bibinfo{journal}{Int. J. Mod. Phys. B} \textbf{\bibinfo{volume}{36}}, \bibinfo{pages}{2250154} (\bibinfo{year}{2022}), \bibinfo{note}{arXiv:2204.07552}, \urlprefix\url{https://www.worldscientific.com/doi/abs/10.1142/S0217979222501545}.

\bibitem[{\citenamefont{Almeida et~al.}(2007)\citenamefont{Almeida, de~Melo, Hor-Meyll, Salles, Walborn, Ribeiro, and Davidovich}}]{Almeida2007EnvironmentEntSuddenDeath}
\bibinfo{author}{\bibfnamefont{M.~P.} \bibnamefont{Almeida}}, \bibinfo{author}{\bibfnamefont{F.}~\bibnamefont{de~Melo}}, \bibinfo{author}{\bibfnamefont{M.}~\bibnamefont{Hor-Meyll}}, \bibinfo{author}{\bibfnamefont{A.}~\bibnamefont{Salles}}, \bibinfo{author}{\bibfnamefont{S.~P.} \bibnamefont{Walborn}}, \bibinfo{author}{\bibfnamefont{P.~H.~S.} \bibnamefont{Ribeiro}}, \bibnamefont{and} \bibinfo{author}{\bibfnamefont{L.}~\bibnamefont{Davidovich}}, \bibinfo{journal}{Science} \textbf{\bibinfo{volume}{316}}, \bibinfo{pages}{579} (\bibinfo{year}{2007}), ISSN \bibinfo{issn}{0036-8075}, \urlprefix\url{http://dx.doi.org/10.1126/science.1139892}.

\bibitem[{\citenamefont{Yu and Eberly}(2009)}]{EntSuddenDeath}
\bibinfo{author}{\bibfnamefont{T.}~\bibnamefont{Yu}} \bibnamefont{and} \bibinfo{author}{\bibfnamefont{J.~H.} \bibnamefont{Eberly}}, \bibinfo{journal}{Science} \textbf{\bibinfo{volume}{323}}, \bibinfo{pages}{598} (\bibinfo{year}{2009}), \urlprefix\url{http://dx.doi.org/10.1126/science.1167343}.

\bibitem[{\citenamefont{Albrecht}(2026)}]{LAQC_Swap}
\bibinfo{author}{\bibfnamefont{H.~L.} \bibnamefont{Albrecht}}, \bibinfo{journal}{Brazilian Journal of Physics} \textbf{\bibinfo{volume}{56}}, \bibinfo{pages}{180} (\bibinfo{year}{2026}), ISSN \bibinfo{issn}{0103-9733}, \urlprefix\url{http://dx.doi.org/10.1007/s13538-026-02104-9}.

\bibitem[{\citenamefont{Werner}(1989)}]{Werner}
\bibinfo{author}{\bibfnamefont{R.}~\bibnamefont{Werner}}, \bibinfo{journal}{Phys. Rev. A} \textbf{\bibinfo{volume}{40}}, \bibinfo{pages}{4277} (\bibinfo{year}{1989}), \urlprefix\url{http://dx.doi.org/10.1103/PhysRevA.40.4277}.

\bibitem[{\citenamefont{Batle and Casas}(2011)}]{Batle2011Nonlocality}
\bibinfo{author}{\bibfnamefont{J.}~\bibnamefont{Batle}} \bibnamefont{and} \bibinfo{author}{\bibfnamefont{M.}~\bibnamefont{Casas}}, \bibinfo{journal}{J. Phys. A: Math. Theor.} \textbf{\bibinfo{volume}{44}}, \bibinfo{pages}{445304} (\bibinfo{year}{2011}), ISSN \bibinfo{issn}{1751-8113}, \urlprefix\url{http://dx.doi.org/10.1088/1751-8113/44/44/445304}.

\bibitem[{\citenamefont{Fan et~al.}(2019)\citenamefont{Fan, Ding, Ming, Yang, Wang, and Ye}}]{Fan2019Inequality}
\bibinfo{author}{\bibfnamefont{X.-G.} \bibnamefont{Fan}}, \bibinfo{author}{\bibfnamefont{Z.-Y.} \bibnamefont{Ding}}, \bibinfo{author}{\bibfnamefont{F.}~\bibnamefont{Ming}}, \bibinfo{author}{\bibfnamefont{H.}~\bibnamefont{Yang}}, \bibinfo{author}{\bibfnamefont{D.}~\bibnamefont{Wang}}, \bibnamefont{and} \bibinfo{author}{\bibfnamefont{L.}~\bibnamefont{Ye}} (\bibinfo{year}{2019}), \urlprefix\url{https://arxiv.org/abs/1909.00346}.

\bibitem[{\citenamefont{Munro et~al.}(2001)\citenamefont{Munro, James, White, and Kwiat}}]{Munro2001Maximizing}
\bibinfo{author}{\bibfnamefont{W.~J.} \bibnamefont{Munro}}, \bibinfo{author}{\bibfnamefont{D.~F.~V.} \bibnamefont{James}}, \bibinfo{author}{\bibfnamefont{A.~G.} \bibnamefont{White}}, \bibnamefont{and} \bibinfo{author}{\bibfnamefont{P.~G.} \bibnamefont{Kwiat}}, \bibinfo{journal}{Phys. Rev. A} \textbf{\bibinfo{volume}{64}} (\bibinfo{year}{2001}), ISSN \bibinfo{issn}{1050-2947}, \urlprefix\url{http://dx.doi.org/10.1103/PhysRevA.64.030302}.

\bibitem[{\citenamefont{Shannon}(1948)}]{Shannon1948Mathematical}
\bibinfo{author}{\bibfnamefont{C.~E.} \bibnamefont{Shannon}}, \bibinfo{journal}{Bell Syst. Tech. J.} \textbf{\bibinfo{volume}{27}}, \bibinfo{pages}{379} (\bibinfo{year}{1948}), ISSN \bibinfo{issn}{0005-8580}, \urlprefix\url{http://dx.doi.org/10.1002/j.1538-7305.1948.tb01338.x}.

\bibitem[{\citenamefont{Girolami and Adesso}(2011)}]{GirolamiAdesso-QD-Xstates}
\bibinfo{author}{\bibfnamefont{D.}~\bibnamefont{Girolami}} \bibnamefont{and} \bibinfo{author}{\bibfnamefont{G.}~\bibnamefont{Adesso}}, \bibinfo{journal}{Phys. Rev. A} \textbf{\bibinfo{volume}{83}}, \bibinfo{pages}{052108} (\bibinfo{year}{2011}), \urlprefix\url{https://link.aps.org/doi/10.1103/PhysRevA.83.052108}.

\bibitem[{\citenamefont{Huang}(2013)}]{Huang-QD-Xstates-WorstCaseScenario}
\bibinfo{author}{\bibfnamefont{Y.}~\bibnamefont{Huang}}, \bibinfo{journal}{Phys. Rev. A} \textbf{\bibinfo{volume}{88}}, \bibinfo{pages}{014302} (\bibinfo{year}{2013}), \urlprefix\url{https://link.aps.org/doi/10.1103/PhysRevA.88.014302}.

\bibitem[{\citenamefont{Luo}(2008)}]{Luo2008QD-BellDiagonal}
\bibinfo{author}{\bibfnamefont{S.}~\bibnamefont{Luo}}, \bibinfo{journal}{Phys. Rev. A} \textbf{\bibinfo{volume}{77}} (\bibinfo{year}{2008}), ISSN \bibinfo{issn}{1050-2947}, \urlprefix\url{http://dx.doi.org/10.1103/PhysRevA.77.042303}.

\bibitem[{\citenamefont{Maldonado-Trapp et~al.}(2015)\citenamefont{Maldonado-Trapp, Hu, and Roa}}]{MaldonadoTrapp-QD}
\bibinfo{author}{\bibfnamefont{A.}~\bibnamefont{Maldonado-Trapp}}, \bibinfo{author}{\bibfnamefont{A.}~\bibnamefont{Hu}}, \bibnamefont{and} \bibinfo{author}{\bibfnamefont{L.}~\bibnamefont{Roa}}, \bibinfo{journal}{Quantum Inf. Process.} \textbf{\bibinfo{volume}{83}}, \bibinfo{pages}{1947} (\bibinfo{year}{2015}), \urlprefix\url{https://doi.org/10.1007/s11128-015-0943-y}.

\bibitem[{\citenamefont{Liao et~al.}(2015)\citenamefont{Liao, Fang, Fang, and Huang}}]{Liao_QD}
\bibinfo{author}{\bibfnamefont{X.}~\bibnamefont{Liao}}, \bibinfo{author}{\bibfnamefont{J.}~\bibnamefont{Fang}}, \bibinfo{author}{\bibfnamefont{M.}~\bibnamefont{Fang}}, \bibnamefont{and} \bibinfo{author}{\bibfnamefont{Z.}~\bibnamefont{Huang}}, \bibinfo{journal}{Int. J. Theor. Phys.} \textbf{\bibinfo{volume}{50}}, \bibinfo{pages}{3340–3349} (\bibinfo{year}{2015}), \urlprefix\url{https://doi.org/10.1007/s10773-011-0759-1}.

\bibitem[{\citenamefont{Fano}(1983)}]{Fano1983}
\bibinfo{author}{\bibfnamefont{U.}~\bibnamefont{Fano}}, \bibinfo{journal}{Rev. Mod. Phys.} \textbf{\bibinfo{volume}{55}}, \bibinfo{pages}{855} (\bibinfo{year}{1983}), \urlprefix\url{https://link.aps.org/doi/10.1103/RevModPhys.55.855}.

\bibitem[{\citenamefont{Rivas et~al.}(2014)\citenamefont{Rivas, Huelga, and Plenio}}]{Plenio2014-qNon-Markov}
\bibinfo{author}{\bibfnamefont{{\' A}.}~\bibnamefont{Rivas}}, \bibinfo{author}{\bibfnamefont{S.~F.} \bibnamefont{Huelga}}, \bibnamefont{and} \bibinfo{author}{\bibfnamefont{M.~B.} \bibnamefont{Plenio}}, \bibinfo{journal}{Reports on Progress in Physics} \textbf{\bibinfo{volume}{77}}, \bibinfo{pages}{094001} (\bibinfo{year}{2014}), ISSN \bibinfo{issn}{0034-4885}, \urlprefix\url{http://dx.doi.org/10.1088/0034-4885/77/9/094001}.

\bibitem[{\citenamefont{Simoen et~al.}(2011)\citenamefont{Simoen, Kaczer, Toledano-Luque, and Claeys}}]{Simoen2011-RTN}
\bibinfo{author}{\bibfnamefont{E.}~\bibnamefont{Simoen}}, \bibinfo{author}{\bibfnamefont{B.}~\bibnamefont{Kaczer}}, \bibinfo{author}{\bibfnamefont{M.}~\bibnamefont{Toledano-Luque}}, \bibnamefont{and} \bibinfo{author}{\bibfnamefont{C.}~\bibnamefont{Claeys}}, \bibinfo{journal}{ECS Transactions} \textbf{\bibinfo{volume}{39}}, \bibinfo{pages}{3} (\bibinfo{year}{2011}), ISSN \bibinfo{issn}{1938-5862}, \urlprefix\url{http://dx.doi.org/10.1149/1.3615171}.

\bibitem[{\citenamefont{Wold et~al.}(2012)\citenamefont{Wold, Brox, Galperin, and Bergli}}]{Wold2012ClassicRTN}
\bibinfo{author}{\bibfnamefont{H.~J.} \bibnamefont{Wold}}, \bibinfo{author}{\bibfnamefont{H.}~\bibnamefont{Brox}}, \bibinfo{author}{\bibfnamefont{Y.~M.} \bibnamefont{Galperin}}, \bibnamefont{and} \bibinfo{author}{\bibfnamefont{J.}~\bibnamefont{Bergli}}, \bibinfo{journal}{Phys. Rev. B} \textbf{\bibinfo{volume}{86}} (\bibinfo{year}{2012}), ISSN \bibinfo{issn}{1098-0121}, \urlprefix\url{http://dx.doi.org/10.1103/PhysRevB.86.205404}.

\bibitem[{\citenamefont{{Mulong Luo} et~al.}(2015)\citenamefont{{Mulong Luo}, {Runsheng Wang}, {Shaofeng Guo}, {Jing Wang}, {Jibin Zou}, and {Ru Huang}}}]{Mulong2015ImpactsOfRTN}
\bibinfo{author}{\bibnamefont{{Mulong Luo}}}, \bibinfo{author}{\bibnamefont{{Runsheng Wang}}}, \bibinfo{author}{\bibnamefont{{Shaofeng Guo}}}, \bibinfo{author}{\bibnamefont{{Jing Wang}}}, \bibinfo{author}{\bibnamefont{{Jibin Zou}}}, \bibnamefont{and} \bibinfo{author}{\bibnamefont{{Ru Huang}}}, \bibinfo{journal}{IEEE Transactions on Electron Devices} \textbf{\bibinfo{volume}{62}}, \bibinfo{pages}{1725} (\bibinfo{year}{2015}), ISSN \bibinfo{issn}{0018-9383}, \urlprefix\url{http://dx.doi.org/10.1109/TED.2014.2368191}.

\bibitem[{\citenamefont{West}(2006)}]{West2006ModelingComplex-RTN}
\bibinfo{author}{\bibfnamefont{B.~J.} \bibnamefont{West}}, \bibinfo{journal}{Complexity} \textbf{\bibinfo{volume}{11}}, \bibinfo{pages}{33} (\bibinfo{year}{2006}), ISSN \bibinfo{issn}{1076-2787}, \urlprefix\url{http://dx.doi.org/10.1002/cplx.20114}.

\bibitem[{\citenamefont{Eberly et~al.}(1984)\citenamefont{Eberly, W{\' o}dkiewicz, and Shore}}]{Eberly1984_RTN-Laser&Atom}
\bibinfo{author}{\bibfnamefont{J.~H.} \bibnamefont{Eberly}}, \bibinfo{author}{\bibfnamefont{K.}~\bibnamefont{W{\' o}dkiewicz}}, \bibnamefont{and} \bibinfo{author}{\bibfnamefont{B.~W.} \bibnamefont{Shore}}, \bibinfo{journal}{Phys. Rev. A} \textbf{\bibinfo{volume}{30}}, \bibinfo{pages}{2381} (\bibinfo{year}{1984}), ISSN \bibinfo{issn}{0556-2791}, \urlprefix\url{http://dx.doi.org/10.1103/PhysRevA.30.2381}.

\bibitem[{\citenamefont{Rivas et~al.}(2010)\citenamefont{Rivas, Huelga, and Plenio}}]{Rivas2010Entang-non_Markov}
\bibinfo{author}{\bibfnamefont{{\' A}.}~\bibnamefont{Rivas}}, \bibinfo{author}{\bibfnamefont{S.~F.} \bibnamefont{Huelga}}, \bibnamefont{and} \bibinfo{author}{\bibfnamefont{M.~B.} \bibnamefont{Plenio}}, \bibinfo{journal}{Physical Review Letters} \textbf{\bibinfo{volume}{105}} (\bibinfo{year}{2010}), ISSN \bibinfo{issn}{0031-9007}, \urlprefix\url{http://dx.doi.org/10.1103/PhysRevLett.105.050403}.

\bibitem[{\citenamefont{Breuer et~al.}(2009)\citenamefont{Breuer, Laine, and Piilo}}]{Breuer2009Measure-non-Markov}
\bibinfo{author}{\bibfnamefont{H.-P.} \bibnamefont{Breuer}}, \bibinfo{author}{\bibfnamefont{E.-M.} \bibnamefont{Laine}}, \bibnamefont{and} \bibinfo{author}{\bibfnamefont{J.}~\bibnamefont{Piilo}}, \bibinfo{journal}{Physical Review Letters} \textbf{\bibinfo{volume}{103}} (\bibinfo{year}{2009}), ISSN \bibinfo{issn}{0031-9007}, \urlprefix\url{http://dx.doi.org/10.1103/PhysRevLett.103.210401}.

\bibitem[{\citenamefont{Zhang and M\o{}lmer}(2020)}]{Zhang2020-OUN_spinB}
\bibinfo{author}{\bibfnamefont{C.}~\bibnamefont{Zhang}} \bibnamefont{and} \bibinfo{author}{\bibfnamefont{K.}~\bibnamefont{M\o{}lmer}}, \bibinfo{journal}{Phys. Rev. A} \textbf{\bibinfo{volume}{102}} (\bibinfo{year}{2020}), ISSN \bibinfo{issn}{2469-9926}, \urlprefix\url{http://dx.doi.org/10.1103/PhysRevA.102.063716}.

\bibitem[{\citenamefont{Turner et~al.}(2022)\citenamefont{Turner, Wu, Li, and Wang}}]{Turner2022-OU_spinB}
\bibinfo{author}{\bibfnamefont{E.}~\bibnamefont{Turner}}, \bibinfo{author}{\bibfnamefont{S.-H.} \bibnamefont{Wu}}, \bibinfo{author}{\bibfnamefont{X.}~\bibnamefont{Li}}, \bibnamefont{and} \bibinfo{author}{\bibfnamefont{H.}~\bibnamefont{Wang}}, \bibinfo{journal}{Phys. Rev. A} \textbf{\bibinfo{volume}{105}} (\bibinfo{year}{2022}), ISSN \bibinfo{issn}{2469-9926}, \urlprefix\url{http://dx.doi.org/10.1103/PhysRevA.105.L010601}.

\bibitem[{\citenamefont{Bell}(1964)}]{Bell1964-OnEPRparadox}
\bibinfo{author}{\bibfnamefont{J.~S.} \bibnamefont{Bell}}, \bibinfo{journal}{Physics Physique Fizika} \textbf{\bibinfo{volume}{1}}, \bibinfo{pages}{195} (\bibinfo{year}{1964}), ISSN \bibinfo{issn}{0554-128X}, \urlprefix\url{http://dx.doi.org/10.1103/PhysicsPhysiqueFizika.1.195}.

\bibitem[{\citenamefont{Clauser et~al.}(1969)\citenamefont{Clauser, Horne, Shimony, and Holt}}]{Clauser1969-CHSH_Ineq}
\bibinfo{author}{\bibfnamefont{J.~F.} \bibnamefont{Clauser}}, \bibinfo{author}{\bibfnamefont{M.~A.} \bibnamefont{Horne}}, \bibinfo{author}{\bibfnamefont{A.}~\bibnamefont{Shimony}}, \bibnamefont{and} \bibinfo{author}{\bibfnamefont{R.~A.} \bibnamefont{Holt}}, \bibinfo{journal}{Physical Review Letters} \textbf{\bibinfo{volume}{23}}, \bibinfo{pages}{880} (\bibinfo{year}{1969}), ISSN \bibinfo{issn}{0031-9007}, \urlprefix\url{http://dx.doi.org/10.1103/PhysRevLett.23.880}.

\bibitem[{\citenamefont{Nunavath et~al.}(2023)\citenamefont{Nunavath, Mishra, and Pathak}}]{Nunavath2023-OpenSysDyn-X}
\bibinfo{author}{\bibfnamefont{N.}~\bibnamefont{Nunavath}}, \bibinfo{author}{\bibfnamefont{S.}~\bibnamefont{Mishra}}, \bibnamefont{and} \bibinfo{author}{\bibfnamefont{A.}~\bibnamefont{Pathak}}, \bibinfo{journal}{Modern Physics Letters A} \textbf{\bibinfo{volume}{38}} (\bibinfo{year}{2023}), ISSN \bibinfo{issn}{0217-7323}, \urlprefix\url{http://dx.doi.org/10.1142/S0217732323500566}.

\end{thebibliography}
\include{biblio-qit}

\end{document}